\pdfoutput=1

\documentclass[useAMS,usenatbib]{mn2e}

\usepackage{graphicx}
\usepackage{amssymb,amsmath}
\usepackage{abbreviations}

\newcommand{\new}[1]{#1}

\newcommand{\rs}{r_{\rm S}}

\title[3D relativistic jet formation]{Three Dimensional Structure of Relativistic Jet Formation}
\author[O. Porth]{O. Porth$^{1,2,3}$\thanks{E-mail:
o.porth@leeds.ac.uk}
\\
$^{1}$Max Planck Institute for Astronomy, K\"onigstuhl 17, 69117 Heidelberg, Germany\\
$^{2}$Center for mathematical Plasma Astrophysics, Department of Mathematics, KU Leuven, Celestijnenlaan 200B, 3001 Leuven, Belgium\\
$^{3}$School of Maths, University of Leeds, LS2 9JT, Great Britain
}
\begin{document}

\date{}

\pagerange{\pageref{firstpage}--\pageref{lastpage}} \pubyear{2002}

\maketitle

\label{firstpage}

\begin{abstract}
Using high resolution adaptive mesh refinement simulations in 3D, we investigate the formation of relativistic jets from rotating magnetospheres.  Here, we focus on the development of non-axisymmetric modes due to internal and external perturbations to the jet. These originate either from injection of perturbations with the flow or from a clumpy external medium.  In the helical field geometry of the accelerating jet, the m=1 to m=5 modes are analyzed and found to saturate at a height of $\sim 20$ inner disk radii. 
We also discuss a means to control artificial amplification of $m=4$ noise in the cartesian simulation geometry.  
Strong perturbations due to an in-homogeneous ambient medium lead to flow configurations with increased magnetic pitch and thus indicate a self-stabilization of the jet formation mechanism.  
\end{abstract}


\section{Introduction}

Relativistic jets from active galactic nuclei (AGN) are host to a multitude of physical processes that lead to emission of radiation with the highest energies in the universe \citep[e.g.][]{bottcher2007}.  
The observed non-thermal emission of high-energy particles is subject to beaming in a highly collimated jet of plasma moving with Lorentz factors in the range $10-40$ \citep{jorstad2005,lister2009}.  
The leading paradigm of jet formation suggests that the acceleration of the bulk flow can be understood in the relativistic magnetohydrodynamic (RMHD) approximation, where the jet is endowed with a large-scale current circuit.  
Current axisymmetric simulation models of jet formation \citep{2007MNRAS.380...51K,2008MNRAS.388..551T,porth2011} have matured to an overall agreement on the underlying process of jet formation.  
However, since nature does not obey axisymmetry, 2D modeling falls short in two important aspects:  
\textit{1.}
The axisymmetric induction equation ($\partial_{\phi}\equiv0$) lacks a mechanism to transform toroidal magnetic field back into poloidal field.  Therefore the toroidal field can only be amplified, leading to a potential overproduction of $B_{\phi}$ and the collimating pinch force. This might cast a shadow of doubt on the validity of results from axisymmetric jet formation simulations.   
\textit{2.}
In order to also study the stability of jets, a three dimensional treatment is necessary.  
It is well known that instability of current carrying plasmas is connected to non-axisymmetric perturbations of the toroidal magnetic field \citep{Bateman:1978}. Among these current driven instabilities, the $m=1$ mode known as the ``kink'' is the most violent one.  It leads to a helical displacement of the flow from the axis of the plasma-cylinder.  
In the context of young stellar objects, the helical kink can yield an explanation for the wiggly structure observed in some stellar jets (e.g. HH 46/47) \citep{todo1993, lery2000}. 
If not brought to a halt by regulating non-linear mechanisms, the exponential instability growth must lead to complete disruption of the jet.  
At the presence of instability, jets can dissipate magnetic energy via reconnection \citep[e.g.][]{2002A&A...391.1141D,2003ApJ...589..893L} and shocks (e.g. \cite{Blandford1979}), leading to heating, acceleration of high-energy particles and radiation.  
As noted by \cite{heinz2000} and \cite{2002A&A...391.1141D}, the dissipation into a fully tangled magnetic field could also promote efficient quasi-thermal acceleration of the bulk flow out of magnetic enthalpy.  
In principle, the current-driven instabilities can be accompanied by other types of instabilities such as the Kelvin-Helmholtz instability (KH) caused by shear between jet and ambient material.  
The KH instability can thus lead to an efficient mixing of the jet and environment.  
However, in the presence of strong magnetic fields, growth of the KH modes is strongly suppressed \citep{keppens1999}. In particular, toroidal fields appear to hinder mixing and thus exert a stabilizing influence \citep{appl1992,Mignone:2010a}.

As pointed out by \cite{mckinney2009}, when the well-known Kruskal-Shafranov (KS) instability criterion 
\begin{align}
\left|\frac{B_{\phi}}{B_{p}}\right|>\frac{2\pi r}{z}
\end{align}
for cylindrical force-free equilibria is applied to relativistic jets, we obtain the result that jets become unstable already at the Alfv\'en point $z_{A}\simeq 10 r_{\rm S}$ (where $B_{\phi}\gtrsim B_{p}$ and $z\gtrsim r$; $r_{\rm S}$ denotes the Schwarzschild radius) - before accelerating to highly relativistic velocities.  
This is in stark contrast to the finding of some AGN FR-II jets propagating unperturbed out to distances of $10^{7} r_{\rm S}$.  

Linear stability analysis in non-relativistic, but otherwise complete MHD was conducted for current-free and current-carrying jets with helical magnetic fields by \cite{appl1992}.  
The stability of \emph{relativistic} MHD jets on the other hand is still not fully addressed in linear analysis and subject of ongoing research. 
Considering linear analysis of \emph{force-free} cylindrical configurations, 
\cite{istomin1994} and \cite{istomin1996} concluded stability of $B_{z}=const.$ jets with respect to axisymmetric and helical perturbations.  
\cite{begelman1998} on the other hand demonstrated the violent instability of the $m=1$ mode which he postulated as a possible solution to the long-standing $\sigma-$problem.  
Also \cite{Lyubarskii:1999} stressed the importance of the $m=1$ mode by considering more realistic configurations with decreasing flux. However, \cite{Lyubarskii:1999} and recently \cite{narayan2009} note the time-dilated slow growth rate of the kink.  
This can lead to a substantial displacement between launching and dissipation scales and thus yield a possible explanation for the extent of the blazar zone of AGN.

\cite{Tomimatsu:2001} could extend the classical KS criterion and demonstrated a stabilizing effect of the relativistic field line rotation.  Their analysis (TKS) yields the simple criterion for instability
\begin{align}
\left|\frac{B_{\phi}}{B_{p}}\right|>\frac{2\pi r}{z}\ \hspace{1cm}{\rm and}\hspace{1cm}
\left|\frac{B_{\phi}}{B_{p}}\right|>\frac{r\Omega}{c}
\end{align}
where $\Omega$ denotes the angular velocity of the field line $\Omega= (v_{\phi} - v_{p} B_{\phi}/B_{p})/r$, 
such that the toroidal field strength also has to overcome the stabilizing electric field.  
The asymptotic relation for relativistic jets $B_{\phi}\simeq -r\Omega/c B_{p}$ suggests marginal stability of the unperturbed flow.\footnote{However, this result should be handled with care, since the TKS criterion strictly only applies in the sub-Alfv\'enic region of the flow; see also the discussion by \cite{mckinney2009}.}
Taken at face value, the TKS criterion thus suggests that relativistic jets are always on the verge of instability with the ultimate fate strongly depending on details of the modeling.  
In more recent studies, several authors investigated the stabilizing influence of environmental effects such as shear \citep[e.g.][]{2007ApJ...662..835M}, external wind with relativistic bulk motion \citep{2003ApJ...583..116H} and sideways expansion \citep{2000ApJ...542..750R}, emphasizing the influence of modeling details for jet and ambient material.   
While there is now growing consensus that the kink instability can also operate in the relativistic regime, to answer whether it can grow indefinitely to finally disrupt the jet or rather saturate, requires a study of the non-linear evolution via numerical simulations \citep[e.g.][]{2009ApJ...700..684M,Mignone:2010a,mizuno2011a}.  
Also \cite{oneill2012a} studied the stability of various local jet configurations in force-free, pressure confined and rotational equilibriua. These authors found pressure confined models corresponding to astrophysical jets furthest from the origin to be the most unstable ones.  

On the other hand we would like to point out that jet stability can best be studied by including the initial acceleration and collimation region of the jet.  
Thereby, self-consistent helical magnetic fields are obtained and the number of ad-hoc assumptions for the jet base can be reduced to a minimum of physically well motivated choices.  
\cite{mckinney2009} were the first to present global 3D simulations of jet formation including a turbulent accretion disk.  In their seminal paper, no significant disruption or dissipation out to scales of $10^{3}\rs$ was observed, however by using comparatively low resolution GRMHD simulations with $256\times128\times32$ cells in a spherical $r\times \theta\times\phi$ grid.
\footnote{Although a fourth order limiter was used, we note here that an accurate study of the higher-order mode growth in the jet requires sufficient actual angular resolution (\cite{mckinney2009} report $20\%$ convergence in the $m=1,2,3$ power when comparing 16 and 32 angular cells).}
Although launching jets directly from the turbulent accretion flow promises high realism, a systematic study of jet instability growth can hardly be performed this way.  
For non-relativistic disk jets, \cite{2003ApJ...582..292O}, \cite{anderson2006}, \cite{2008A&A...492..621M} and \cite{staff2010} followed an alternative approach by treating the rotating magnetosphere as a fixed-in time injection boundary.  
As a natural generalization of previous axisymmetric work, the boundary can be used to model the corona of a Keplerian accretion disk.  
Also we will follow this strategy for the case of relativistic jets, as it allows to systematically control both fixed in time and time variable injection conditions.

In this paper, we show our first results concerning the stability of relativistic jets near the launching region.  We will discuss 3D RMHD simulations exposed to non-axisymmetric perturbations triggered by the accretion disk and due to jet-cloud interactions.  To complement previous considerations of axisymmetric (2.5D) jet formation outlined in \cite{2010ApJ...709.1100P} (PF I) and \cite{porth2011} (PF II), we focus mainly on non-axisymmetric features within the jet.

\section{Model Setup}

The jet is launched from a rotating inlet resembling the corona of an accretion disk similar to PF I.  
The initially purely poloidal magnetic field is transformed into a helical shape giving rise to a global electric current system which accelerates and collimates the flow into a jet.  
\new{Since the disk enters our description essentially as a boundary (allowing for outgoing Alfv\'en and fast waves) we do not investigate the back reaction from the jet to the disk.}
Due to the increased complexity of the three dimensional treatment compared to previous 2.5D work, we neglect stratification caused by a gravitational source term, since the flow is accelerated above escape velocity very near the disk and sonic surface (see also PF I).  
Note that in the following, the axis of symmetry is denoted as the cartesian 'y-axis' (not the 'z-axis' as usual).

\subsection{Initial and Boundary Conditions}\label{sec:3dboundary}

\begin{figure}
\begin{center}
\includegraphics[width=0.48\textwidth]{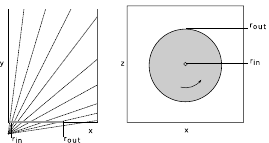}
\caption{Sketch of the computational domain in the x-y plane \textit{(left)} and the x-z plane \textit{(right)}.  Within $r_{\rm out}$, rotating injection conditions are applied. The monopolar magnetic field is indicated by dashed lines in the xy plane and has its ``source'' outside of the domain at $(x,y,z) = (0,-4,0)$.  }
\label{fig:domain}
\end{center}
\end{figure}

We initialize the domain with constant values for density and pressure $\rho\equiv\rho_{0}=0.01$, $p\equiv p_{0}$, threaded by an initial poloidal magnetic field of monopolar shape with the ``source'' outside of the domain at $y=-4$.  The initial pressure follows from the plasma $\beta$ parameter at the inner disk edge $r=1$ to $p_{0}=\beta/2 B_{1}^{2}$, where $B_{1}$ denotes the corresponding poloidal magnetic field strength at $(r,y)=(1,0)$ and we set to $B_{1}=1$ for convenience.  A sketch of the computational domain is shown in figure \ref{fig:domain}.  

To avoid the poorly determined states in the corners of the domain where disk-boundary and outflow boundary would meet, we truncate the disk inlet at $r_{\rm out}$.  
Not truncating the disk introduces additional $m=4$ noise triggered predominantly at the corners of the domain.  
At the injection boundary within $r<r_{\rm out}$ we assign boundary conditions for a rotating magnetosphere, injecting a sonic flow aligned with the poloidal field-lines.  
To provide the axisymmetric boundary constraints at the bottom plane, we first transform to cylindrical coordinates around the jet axis in the middle of the cartesian domain, compute the required quantities and subsequently transform back to the cartesian domain to assign the updated boundary conditions.   
We specify fixed-in time axisymmetric profiles for the five quantities $p_{\rm d}\equiv p_{0}, \rho_{\rm d}\equiv 100\rho_{0}=1, v_{\phi}=r \omega(r), v_{p}\equiv c_{\rm s}, E_{\phi}\equiv0$ to account for the downstream Alfv\'en and fast magnetosonic waves leaving the domain across the boundary.  In addition, the magnetic flux $B_{y}$ is fixed to the initial profile.  Since the evolution of $B_{y}$ is already suppressed via the choice of $E_{\phi}\equiv0$, this does not represent an additional boundary constraint.  The radial field component is obtained by a field extrapolation satisfying $j_{\phi}\equiv0$ of the domain values $B_{r}$, given the fixed profile of the vertical field component $B_{y}$.  
The toroidal field $B_{\phi}$ follows from the domain via zero gradient extrapolation to yield also $j_{r}=0$.  
This way, the boundary current is under control and no spurious current sheets are allowed to arise at the bottom plane injection boundary.  
For the rotation profile, we adopt 
\begin{align}
\omega(r) = 0.5 c/r_{\rm in} \left\{\begin{array}{lll}
1 &; & r<r_{\rm in}\\
\left(\frac{r}{r_{\rm in}}\right)^{-3/2} &; & 1\le r <r_{\rm out}
\end{array}
\right.
\end{align}
representing an inner solid-body rotation law and an outer Keplerian profile within $r_{\rm in}=1 $ as in \cite{2008MNRAS.388..551T}. Beyond $r_{\rm out}$, we simply freeze the initial conditions in the boundary.  
 
\new{
We should note here that in a general relativistic treatment, frame dragging of a monopolar magnetosphere leads to an approximate solid-body field-line rotation with $\Omega \simeq 0.5\, \Omega_{\rm H}$, where $\Omega_{\rm H}$ denotes the outer horizon angular velocity} \citep{1977MNRAS.179..433B, komissarov2001}.  
\new{In the extreme Kerr case, this would yield $\Omega \simeq 0.25\, c/r_{\rm in}$ hence we are even exaggerating the central field line rotation and thus the putative Blandford \& Znajek power output.  
Naively, one could thus expect a fast axial jet to arise in our simulations.  However, since the mass influx into the domain is in fact increasing as one approaches the origin (given constant \emph{poloidal} disk- velocity and density) and due to the necessity of vanishing Poynting flux on the axis, our simulations will not produce a high Lorentz factor spine.   This would require an accurate modeling of the mass flux in the plunging region using a GRMHD approach which we must defer to future work.  
}

For the set of parameters shown here, $\beta=0.01, v_{\phi,\rm r=1}=0.5c$ and $\rho_{d}=1$, the injection speed follows as $c_{\rm s}\simeq0.1 c$ and hence the maximal \cite{1969ApJ...158..727M} magnetization parameter can be written 
\begin{align}
\sigma_{\rm M} \simeq \frac{v_{\phi}^{2} B_{1}^{2}}{4\pi\Gamma\rho_{\rm d} c_{\rm s}c^{3}}. 
\end{align}
It becomes $\sigma_{\rm M} = 2.5$ giving an upper limit of the Lorentz factor to be obtained by the flow.  Higher values of $\sigma_{\rm M}$ can be achieved by reducing the disk density or the plasma-$\beta$ as the magnetization depends on these parameters according to $\sigma_{\rm M}\propto \rho^{-1/2} \beta^{-1/2}$.

Due to the accumulation of numerical errors and decreased resolution compared to previous axisymmteric studies, steep gradients of the rotation law at the inner disk edge must be avoided if no special treatment for the velocity profile is given.\footnote{See also the discussion in Appendix 3 of \cite{2003ApJ...582..292O}.}  It is also for similar stability reasons that we specify $\omega$ as a boundary constraint instead of the field rotation law $\Omega$ as customary.  

To close the set of ideal special relativistic magnetohydrodynamic equations (RMHD), we employ the equation of state proposed by \citet{meliani2004}, \new{approximating an ideal one-component monoatomic} \citet{synge1957} \new{gas (see also equation 6 of }\citet{Keppens2012718}).  

Additional passive tracer scalars $\iota_{1}-\iota_{3}$ are advected with the flow for the purposes of refinement and post-processing.  

\subsection{Numerical Grid Setup}

We perform simulations in a cartesian adaptive grid using MPI-AMRVAC \citep{Keppens2012718}.   
With the adaptive mesh refinement in a cartesian domain, a uniformly high resolution for all regions of interest is obtained without the directional biases present in stretched grid simulations \citep[e.g.][]{2008A&A...492..621M, Mignone:2010a}, especially, a high resolution can be achieved also at the jet head which will prove necessary to resolve the helically displaced tip of the jet.  
The cartesian discretization and quadrantal symmetry of the grid can however introduce notable noise leading to a ``pumping'' of the $m=4$ mode.  We comment on the measures taken to analyze and circumvent such spurious effects further in section \ref{sec:modeanalysis}.  

The largest domain size considered in the following extends over $x\in [-32,32]$, $y\in[0,128]$ and $z\in [-32,32]$ in units of the inner disk radius.  A base resolution of $n_{x}\times n_{y} \times n_{z} = 48\times 96 \times 48$ cells is chosen, adaptively refined by four additional grid levels (three for the smaller domain case).  Thus 12 grid cells per inner disk radius are achieved, totaling in 240 cells across the entire jet inlet radius.  To our knowledge this high resolution is unprecedented in 3D jet formation simulations.  The effective resolution for the large domain is $768\times1536\times768=9.06\times10^{8}$ cells and we observe a grid filling with $\simeq3\times10^{8}$ cells at the termination time.  

\subsubsection{Refinement Strategy}
Refinement to the highest level is enforced for the disk inlet $r<r_{\rm out}; y<1$ and for the region around the axis $x,z<1;y<10$ to resolve the steep gradients of the monopole field.  
The jet is refined based on the scheme proposed by \cite{lohner1987} with weighted contributions of density, Lorentz factor and magnetic field strength.  

As additional criterion, the initial domain is coarsened to the lowest level until the jet comes within reach.  This decision is based on the Lorentz factor ($\Gamma>1.05$) and the passive tracer scalars described earlier.  
The criterion in terms of the Lorentz-factor is required in order to avoid accidental coarsening of the torsional Alfv\'{e}n wave launched at the switch on of disk-rotation; it does not transport jet-inlet material and is thus unaffected by the tracers injected along the jet.  
With this strategy, the fast jet is always resolved appropriately, while the enforced coarsening of the initial domain significantly speeds up the simulations.
Especially when the domain is initialized with a non-homogeneous medium, the coarsening-strategy is crucial.  

\subsection{Perturbations}

In order to investigate the behavior upon instability and to break the quandrantal symmetry of the grid, the initially axisymmetric setup needs to be distorted by non-axisymmetric perturbations.  
The physical origin of the perturbation can be found 
in a non-axisymmetric evolution of the accretion disk, for example due to orbit of vortices and quasi-periodic oscillations \citep{van-der-klis1985}, or in a non-homogeneous external medium.  
In active galactic nuclei, the presence of such a medium within the central region is well established as the broad line region that is possibly comprised of clouds orbiting the black hole at high velocity ($\sim 0.1 c$) \citep[e.g.][]{davidson1979,araudo2010} or in manifestation of a clumpy torus surrounding the accretion disk \citep[e.g.][]{dullemond2005}. 
The eventual collision of such clouds with a relativistic jet was already proposed by \cite{Blandford1979} as a mechanism to explain transient features in compact radio sources.  The scenario of jet-cloud collision is thus of interest not only in respect to jet stability, but also concerning the triggering of particle acceleration at the shock surface.   We take the presence of a clumpy ambient medium as a motivation for our second setup of jet perturbations, where we model how the emerging jet propagates through such a density structure.  

\subsubsection{Mode Injection}
To model non-axisymmetric features of the accretion disk, we employ perturbations to the rotation velocity $\omega+\Delta\omega$ following a mode decomposition.  This method was already successfully applied by \cite{rossi2008} and \cite{Mignone:2010a}.  
In particular, we adopt
\begin{align}
\begin{split}
\Delta \omega(r,t) = \\
\frac{\epsilon_{\omega} \omega_{0}(r)}{32}\sum_{m=0}^{3}\sum_{l=1}^{8} \cos(m\phi+\omega(l)\omega_{0}(r) t+\phi_0(l))
\end{split}
\end{align}
to obtain modes with sub- and super-Keplerian  frequencies $\omega(n)\in\{0.5,1,2,3,0.03,0.06,0.12,0.25\}$ featuring a random phase offset $\phi_{0}(l)$.  The maximum amplitude of the perturbation is set to $\epsilon_{\omega}=2\%$.

\subsubsection{Clumpy medium}
A static clumpy medium is modeled as density perturbation by prescribing the size spectrum in Fourier space and subsequent transformation of the random-phased Fourier coefficients to real space.  
To obtain a particular cloud size, the following spectrum is adopted:
\begin{align}
f(k)=  k_{\rm min}^{-s-1}\frac{s+1}{2s}\left\{\begin{array}{clc}
    k^{s}&; & k\ge k_{\rm min} \\ 
    k_{\rm min}^{s}&; & k<k_{\rm min} \\ 
  \end{array}\right.
\end{align}
where $k_{\rm min}$ relates to the cloud size $\lambda$ and domain size via $\lambda={\rm max}(\Delta x, \Delta y, \Delta z)/k_{\rm min}$.  The highest frequency $k_{\rm max}$ is limited by the base-level resolution of the grid to $k_{\rm max}<0.5 \min(n_{x},n_{y},n_{z})$ as we do not allow refinement on the initial condition.  We adopt a slope of the cloud size function $s=5/3$.  In the following, the maximal cloud density is set to $100$ times the ambient density $\rho_{0}$.  

\section{Results and Discussion}

We now describe the simulations performed and the analysis of non-axisymmetric features.  The non-linear temporal and spatial evolution of angular modes is discussed and we show evidence for jet self-stabilization in the launching region.  

\subsection{Overview of the Simulations}

\begin{figure}
\begin{center}
\includegraphics[width=0.45\textwidth]{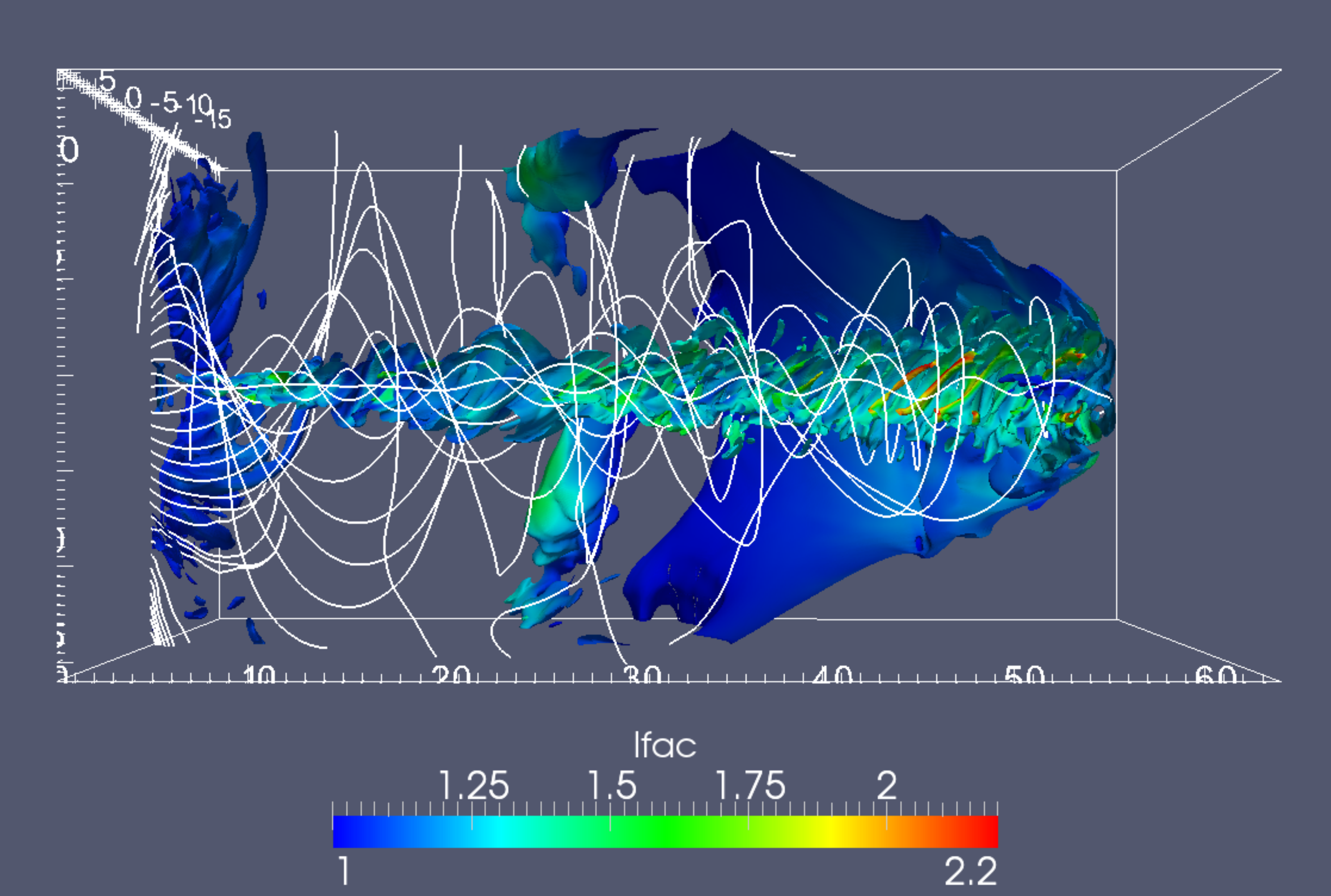}
\caption{Rendering of the reference simulation, showing cut open pressure isocontour colored according to Lorentz factor with magnetic field lines (run M3D). }
\label{fig:fancyplots}
\end{center}
\end{figure}

Table \ref{tab:3D} summarizes our parameter runs and Figure \ref{fig:fancyplots} shows a rendering of our reference unperturbed solution labeled as M3D.

\begin{table}
\caption{Parameter summary of the 3D simulations. Simulation names indicate box size (L,M) and the nature of perturbation (m - mode injection, d - density perturbations). \label{tab:3D}}
\centering
\tiny
\begin{tabular}{lcccccc|c}
\hline\hline
ID & $\beta$ &$\epsilon_{\omega}$ & $r_{\rm out}$ & $\lambda$ & Domain & levels & $T_{\rm end}$\\
\hline
L3D & 0.01 & 0 & 20&   - & $64\times128\times64$ & 5 & 168\\
L3Dm& 0.01 & 0.02 & 20 &- & $64\times128\times64$ & 5 & 168\\
M3D& 0.01 & 0 & 10 & - & $32\times64\times32$& 4 & 103\\
M3Dd& 0.01 & 0 & 10 & 16 & $32\times64\times32$& 4 & 189\\
M3Dmd& 0.01 & 0.02 & 10 & 16 & $32\times64\times32$& 4 & 222\\
\end{tabular}
\end{table}

\begin{figure}
\begin{center}
\includegraphics[width=0.23\textwidth]{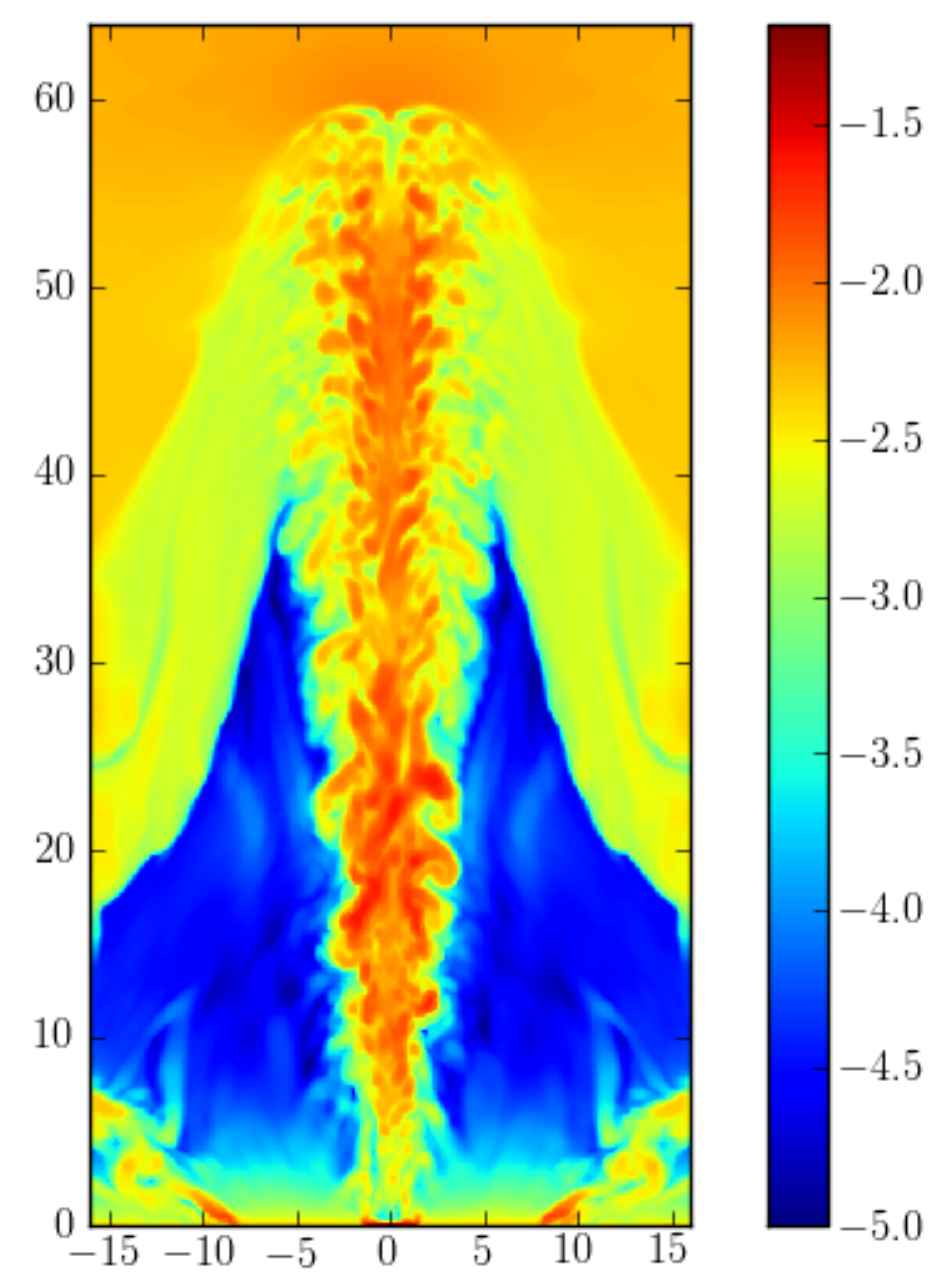}
\includegraphics[width=0.23\textwidth]{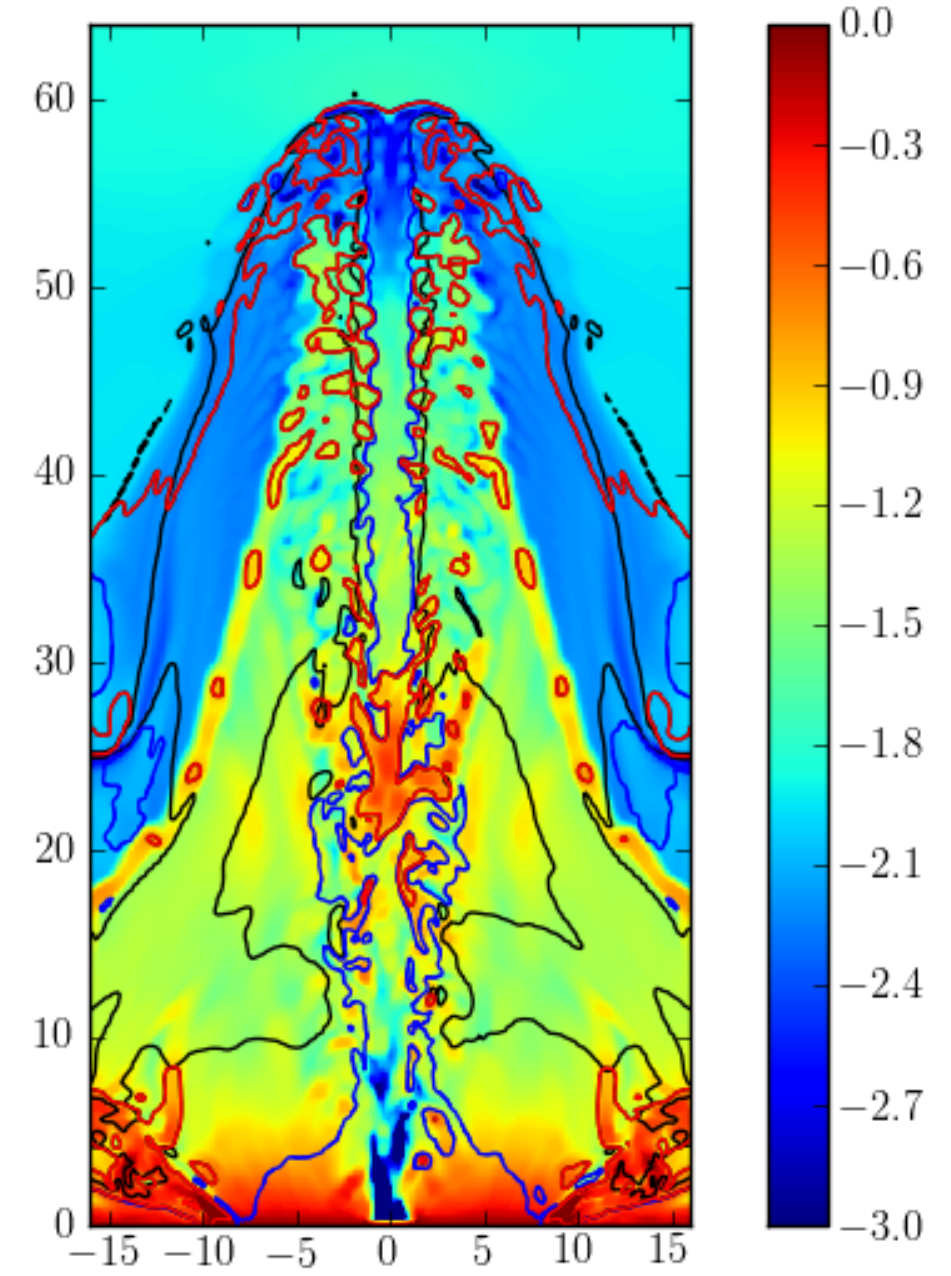}
\includegraphics[width=0.23\textwidth]{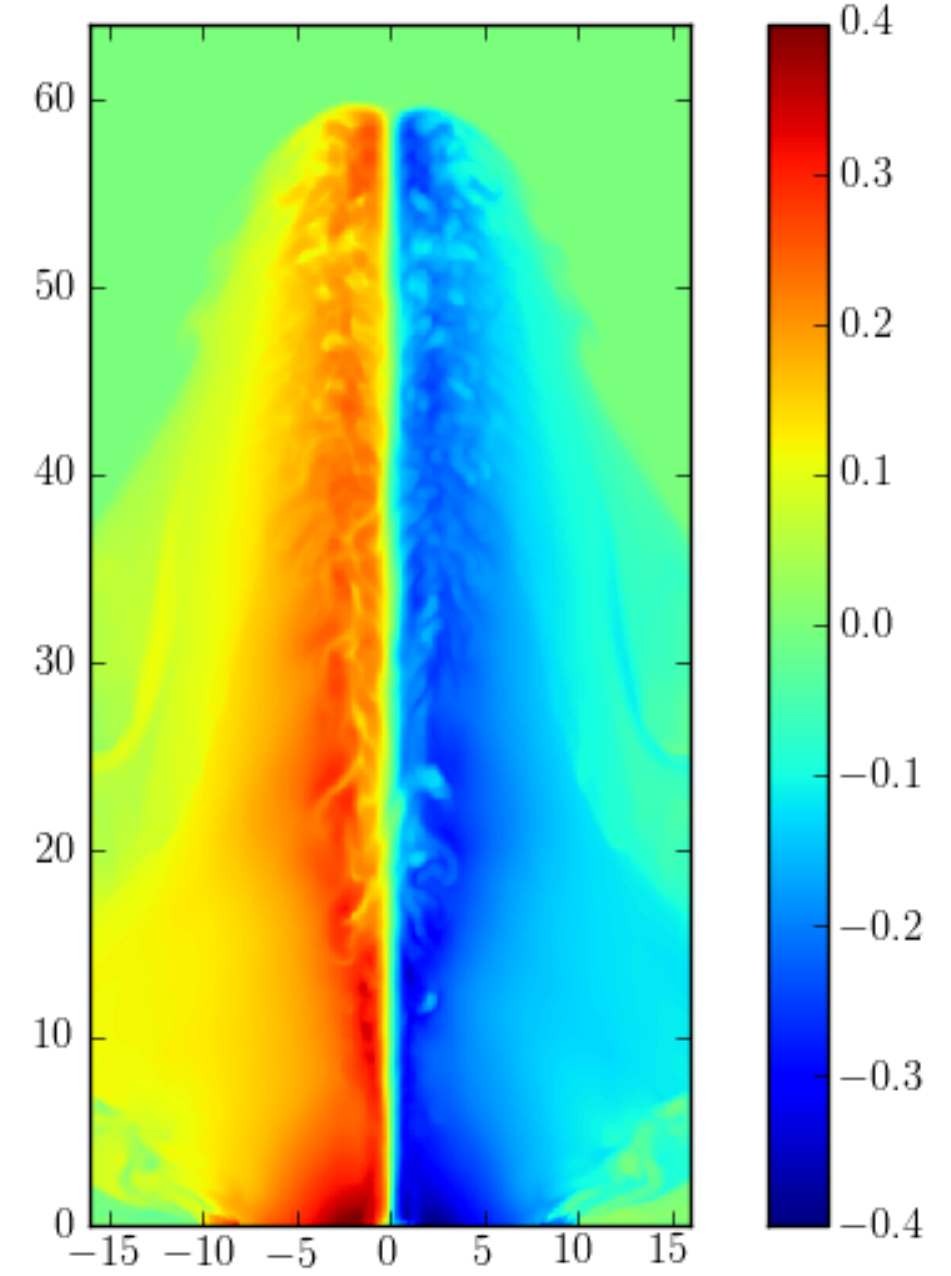}
\includegraphics[width=0.23\textwidth]{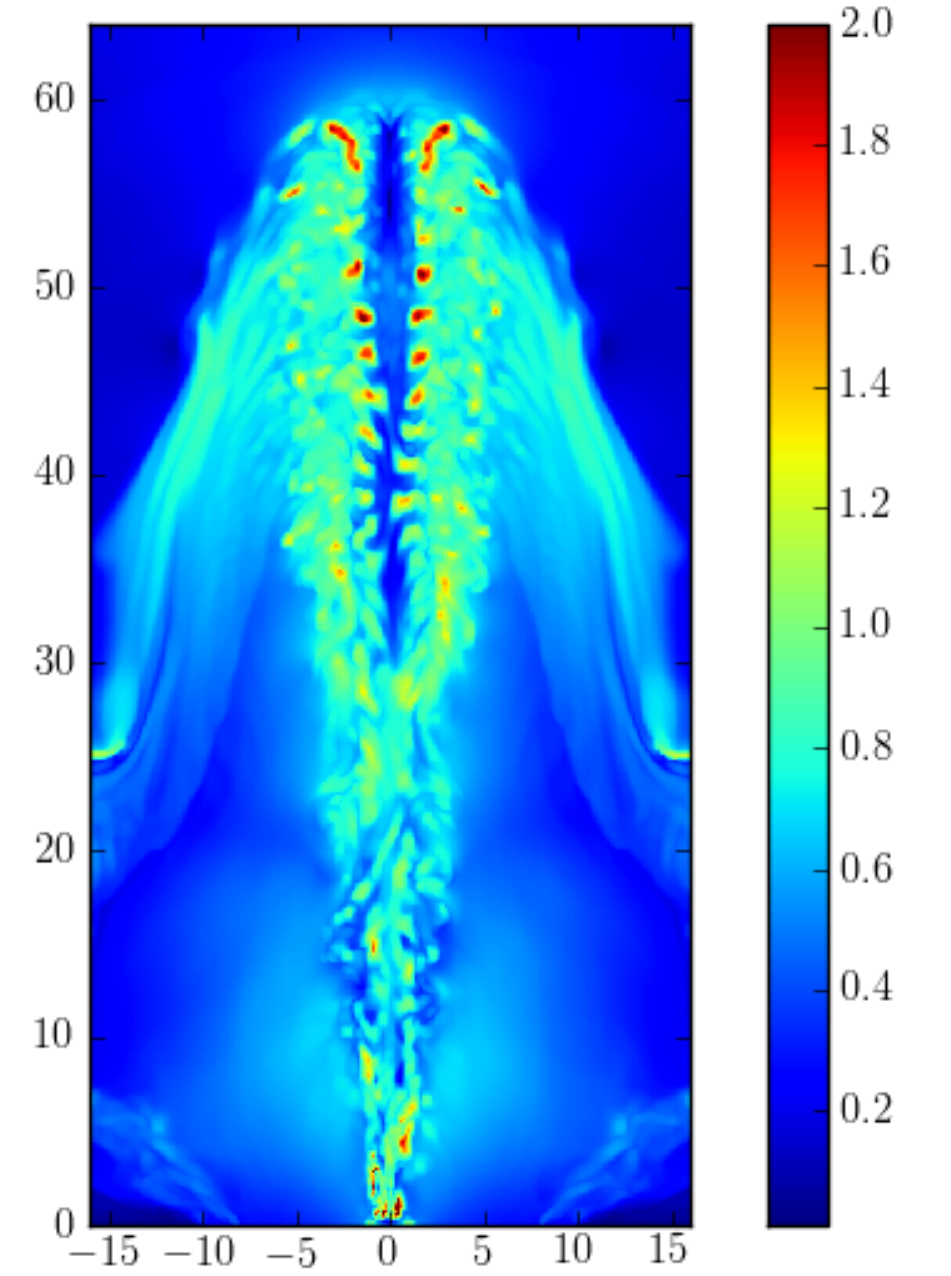}
\caption{Slice along the jet in simulation M3D at $t=103$ $z=0$.  Shown are (left to right, top to bottom): Thermal pressure (log-scale); co-moving density (log-scale); (toroidal) magnetic field strength across the plane and magnitude of four velocity $u=\Gamma v$.  \new{Critical points are indicated in the plot of density, where we show the (critical) Alf\'en surface (blue), the light cylinder (black) and the fast magnetosonic surface (red) according to the definitions of} \citet{Camenzind1986}.  }
\label{fig:valuecomp1}
\end{center}
\end{figure}

Various flow quantities in the $z=0$ plane containing the jet ``axis'' are shown in Figure \ref{fig:valuecomp1}.  \new{Here, we also illustrate the momentary run of the critical surfaces as in PF I.  Note that these are strictly only valid under the assumption of stationarity and axisymmetry.  Compared to PF I, our new simulations are not evolved long enough to adopt the near-stationary state and still reflect the initial transient phase of the sudden spin up of the disk.  
None the less, the flow transcends the critical Alfv\'en surface (blue contour) and becomes super-luminal (black contour) but stays sub magnetofast (red contour) within the domain.}  

At this scale, the Lorentz factor of the disk wind is still moderate ($\Gamma<2$), it rises above $\Gamma\simeq2$ only in regions localized adjacent to the high pressure backbone.

\subsection{Mode Analysis}\label{sec:modeanalysis}

\begin{figure*}
\begin{center}
\includegraphics[width=0.95\textwidth]{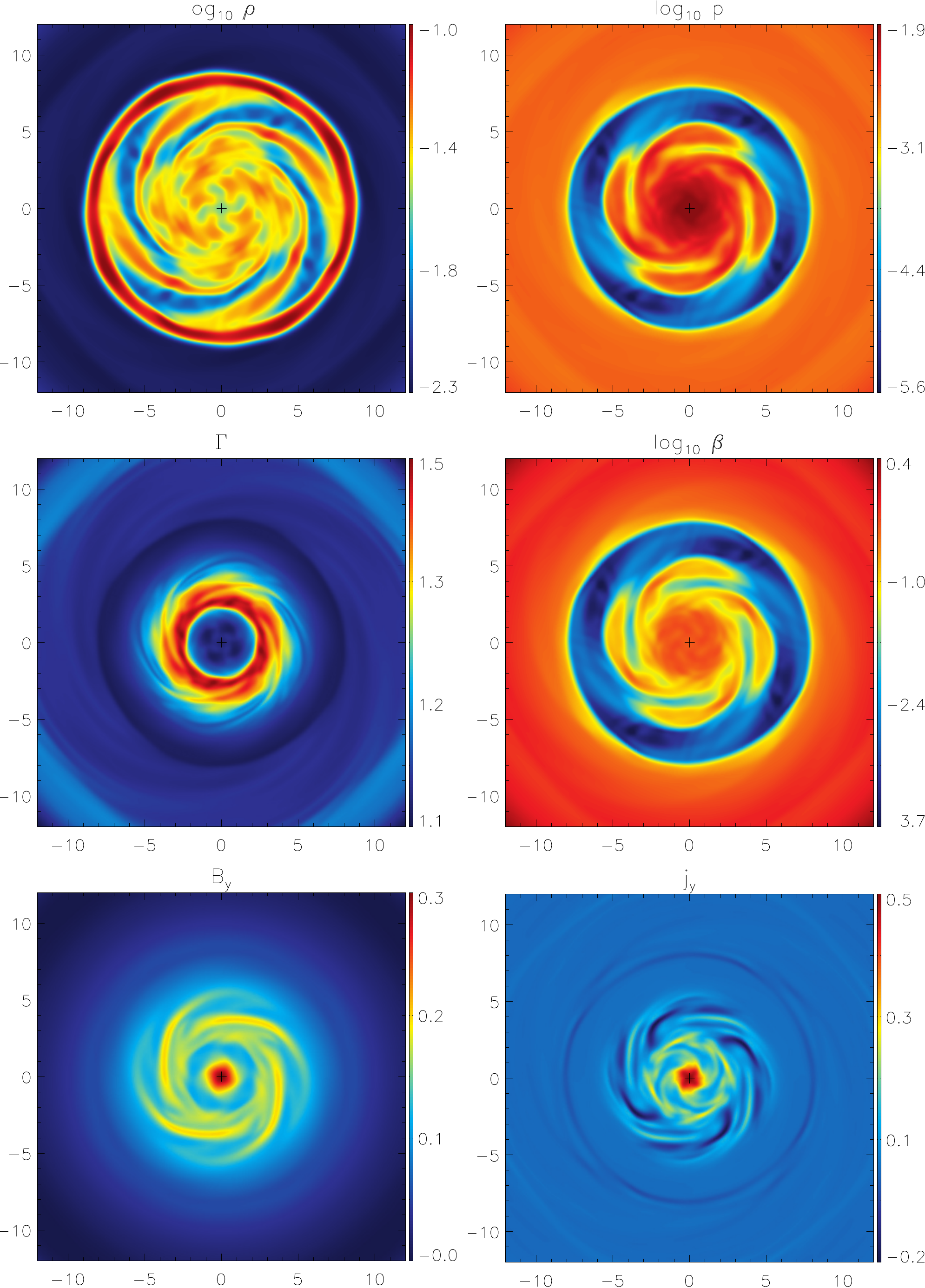}
\caption{Two-dimensional slices in the $x-z$ plane at $y=32$ for $t=100$ in the unperturbed run L3D.  The center is marked by $``+''$.  }
\label{fig:nomodes}
\end{center}
\end{figure*}

\begin{figure*}
\begin{center}
\includegraphics[width=0.95\textwidth]{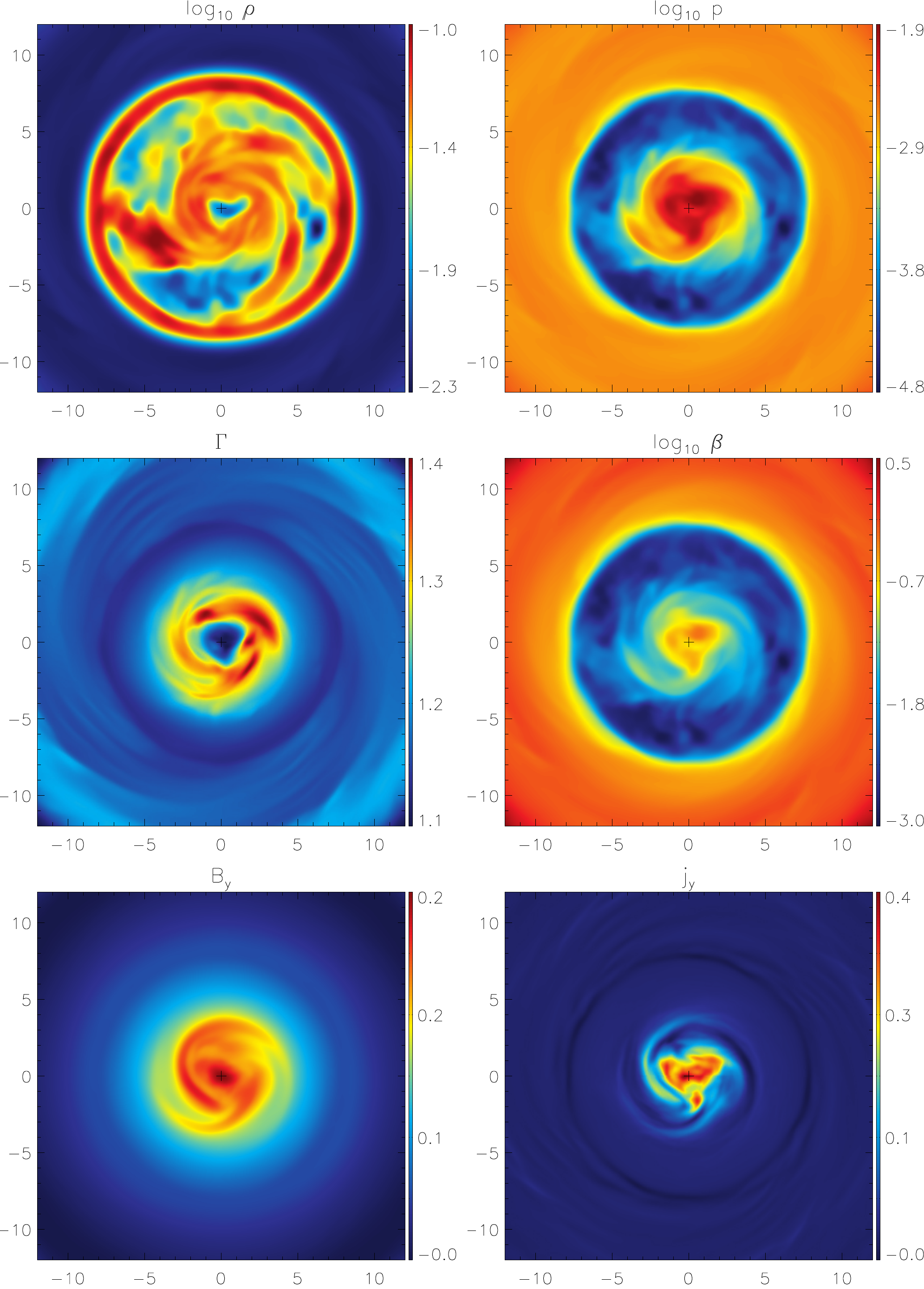}
\caption{As Figure \ref{fig:nomodes} but for the run featuring mode-injection, L3Dm.  }
\label{fig:modes}
\end{center}
\end{figure*}

We now first evaluate the impact of artificial $m=4$ pumping due to the quadrantal symmetry and quantify the growth of non-axisymmetric modes within the jet formation region.  
To first give an impression of the azimuthal variations, a qualitative comparison of the unperturbed simulation L3D with run L3Dm is shown in slices of selected quantities midway across the jet in Figures \ref{fig:nomodes} and \ref{fig:modes}.  
When no measure of perturbing the quadrantal symmetry is taken, we observe multiples of the $m=4$ mode in all flow quantities.  Higher order modes are most apparent in the density $\rho$ and vertical current density $j_{y}$.  In the mode injected slices on the other hand, we observe a slight dominance of the $m=3$ mode.  

\begin{figure}
\begin{center}
\includegraphics[width=0.23\textwidth]{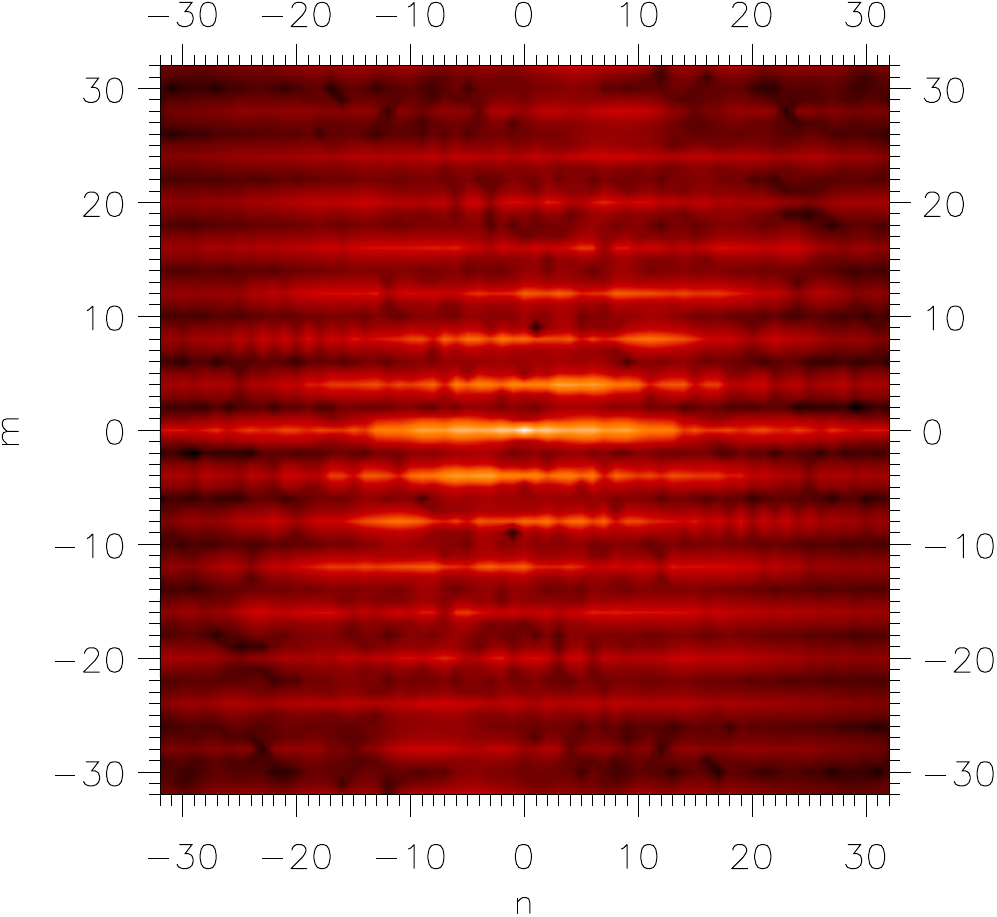}
\includegraphics[width=0.23\textwidth]{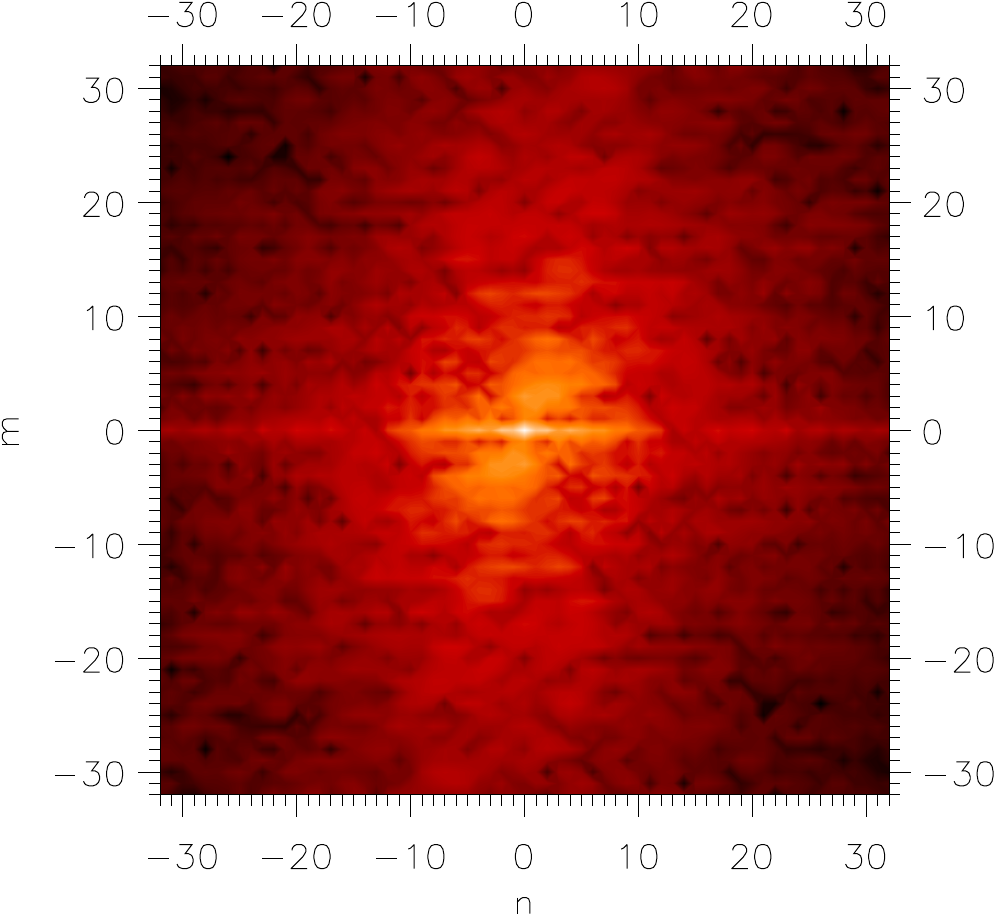}
\vspace{0.4cm}
\includegraphics[width=0.46\textwidth]{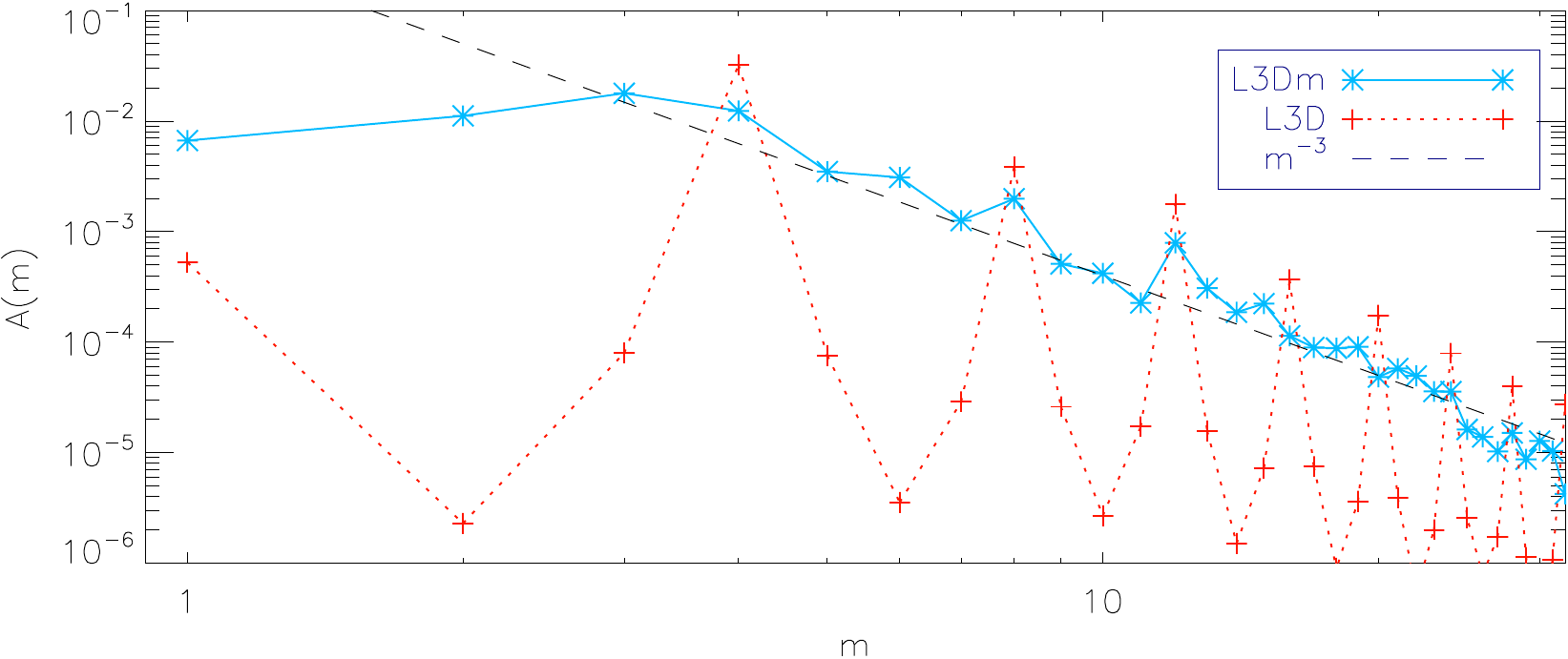}
\caption{\textit{Top:} Radial (n) and angular (m) Fourier amplitudes of density across $y=32$ at $t=100$ for the unperturbed case L3D \textit{(left)} and for the perturbed case L3Dm \textit{(right) }.  
\textit{Bottom:} Cumulative modes for the two cases.  To guide the eye, we show the empirical mode-decay following the power-law $m^{-3}$.  }
\label{fig:powerspecs}
\end{center}
\end{figure}

To quantify the growth of non-axisymmetric modes within the jet, we calculate the fast Fourier transform of the variables on the slice in cylindrical $(r,\phi)$ representation.  
For this purpose, we re-grid the unstructured slice data $x\in[-12,12]$, $z\in[-12,12]$ containing the jet spine to a uniform grid before transformation to the cylindrical coordinates $r\in[0,12]$, $\phi\in[0,2\pi]$.  Thereafter, the Fourier-transformation of the $(r,\phi)$-plane 
\begin{align}
\begin{split}
\tilde{f}(n,m)&=\sum_{n_{r}=-N_{r}}^{N_{r}}\sum_{n_{\phi}=-N_{\phi}}^{N_{\phi}}f(n_{r},n_{\phi})\\
&\times \exp{\left(-2\pi i \left(m \frac{n_{\phi}}{N_{\phi}}+ n \frac{n_{r}}{N_{r}}\right)\right)}
\end{split}
\end{align}
is executed which yields the radial $(n)$ and angular $(m)$ Fourier amplitudes of the input scalar via $A(n,m)\equiv|\tilde{f}(m,n)|^{2}$.  To quantify fluctuations of the angular part alone, we define the normalized cumulative Fourier amplitudes
\begin{align}
A(m)\equiv\frac{1}{|\tilde{f}(0,0)|^{2}}\sum_{n=-N_{n}+1}^{N_{n}-1}|\tilde{f}(n,m)|^{2}
\end{align}
which measures the fluctuations with angular frequency $m$ in relation to the squared mean of the scalar $f$ expressed by ${|\tilde{f}(0,0)|^{2}}$.  
The Fourier amplitude planes of the density fluctuations of Figure \ref{fig:nomodes} and \ref{fig:modes} are shown in Figure \ref{fig:powerspecs}.  
The $m=4$ pollution of the unperturbed run is clearly visible, mode injection on the other hand can be used to get rid of this effect almost entirely as shown in the lower panel of Figure \ref{fig:powerspecs}.

\subsubsection{Temporal Evolution}

\begin{figure}
\begin{center}
\includegraphics[width=0.46\textwidth]{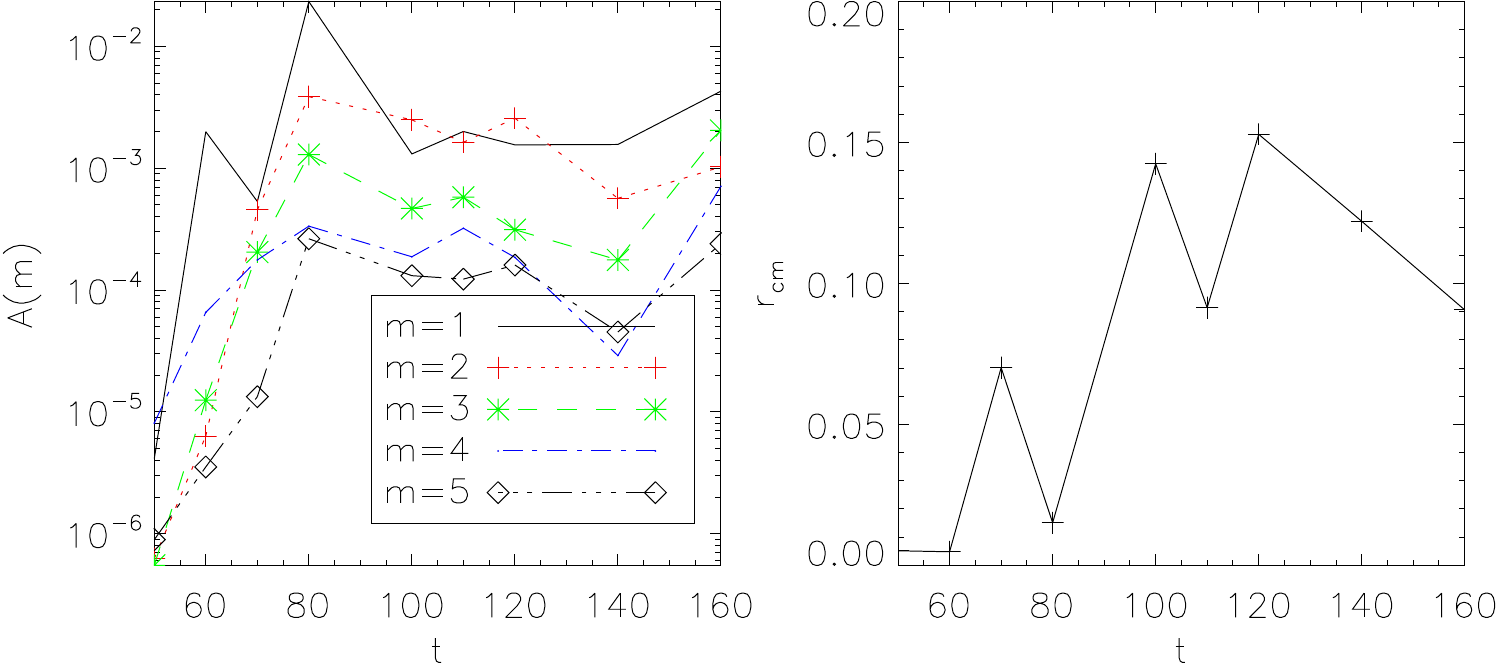}
\caption{\textit{Left:} Mode growth of $B_{y}$ at $y=32$ in simulation L3Dm.  After initial exponential rise, the modes tend to saturate.  Due to grid-noise, the $m=4$ mode is initially comparable to the dominant $m=1$ mode.  \textit{Right:} Barycenter motion on the $y=32$ slice.  }
\label{fig:ampvstime}
\end{center}
\end{figure}

To quantify the temporal growth of the modes, we calculate the Fourier amplitudes at the $y=32$ slice in run L3Dm for various snapshots.  
At this altitude, the magnetic ``backbone'' $B_{y}$ becomes distorted at $t>80$ with dominating $m=1$ and $m=2$ modes, however the amplitudes grow no further.  
The time evolution of the $A(m)$ function is shown in the left panel of Figure \ref{fig:ampvstime}.  After an exponential rise where the growth time in $m=1$ is shortest, followed by the $m=4$ mode, the perturbations saturate at $t\simeq80$ and subsequently fluctuate about a mean value.  Mostly, the amplitudes are ordered according to $A(m)>A(m+1)$ although the $m=2$ occasionally surpasses the $m=1$ contribution.

\subsubsection{Spatial Evolution}

\begin{figure}
\begin{center}
\includegraphics[width=0.46\textwidth]{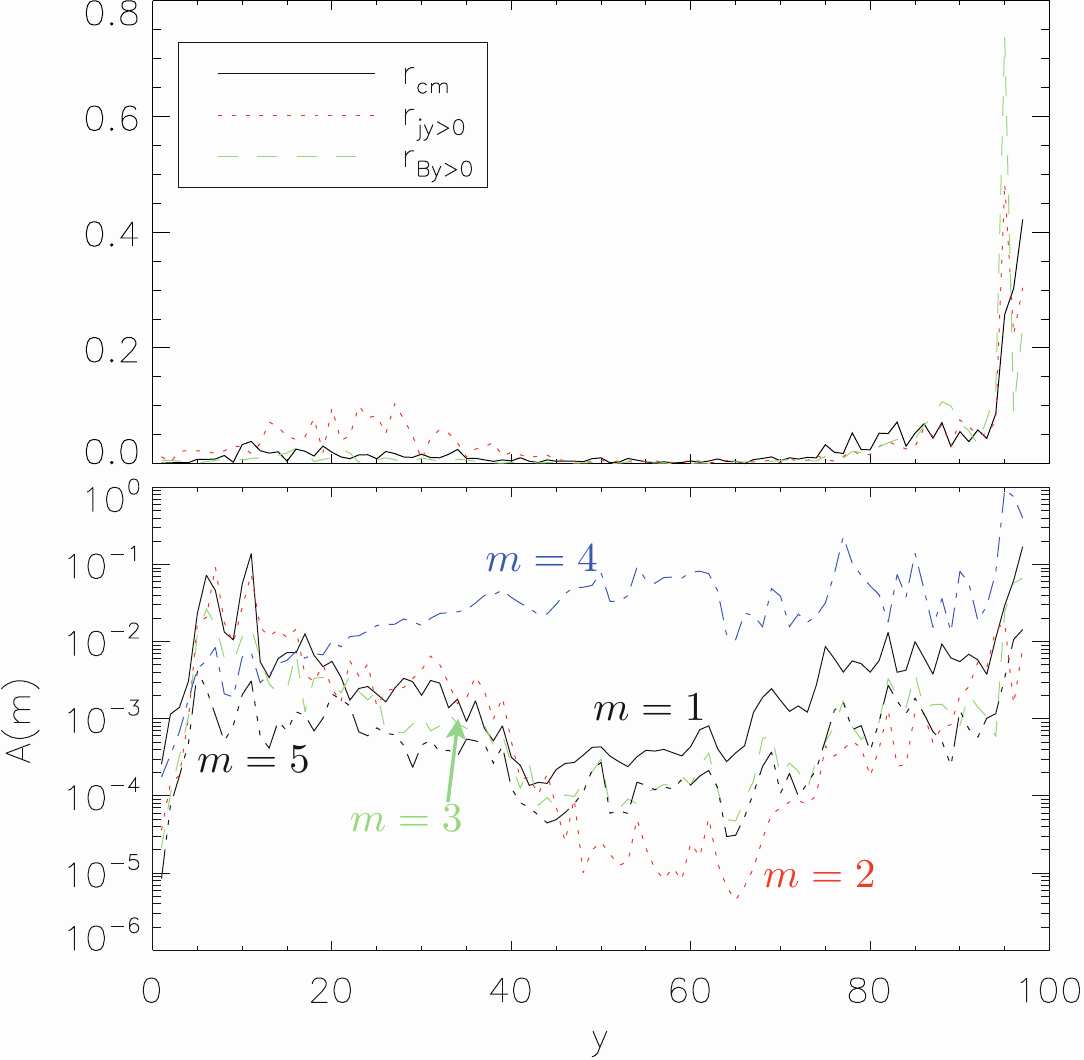}
\caption{Instantaneous barycenter displacements in units of inner disk radius and cumulative Fourier amplitudes of density $\rho$, all along the unperturbed jet L3D at time $t=168$.  The jet base is at left.  }
\label{fig:barycenters}
\end{center}
\end{figure}

\begin{figure}
\begin{center}
\includegraphics[width=0.46\textwidth]{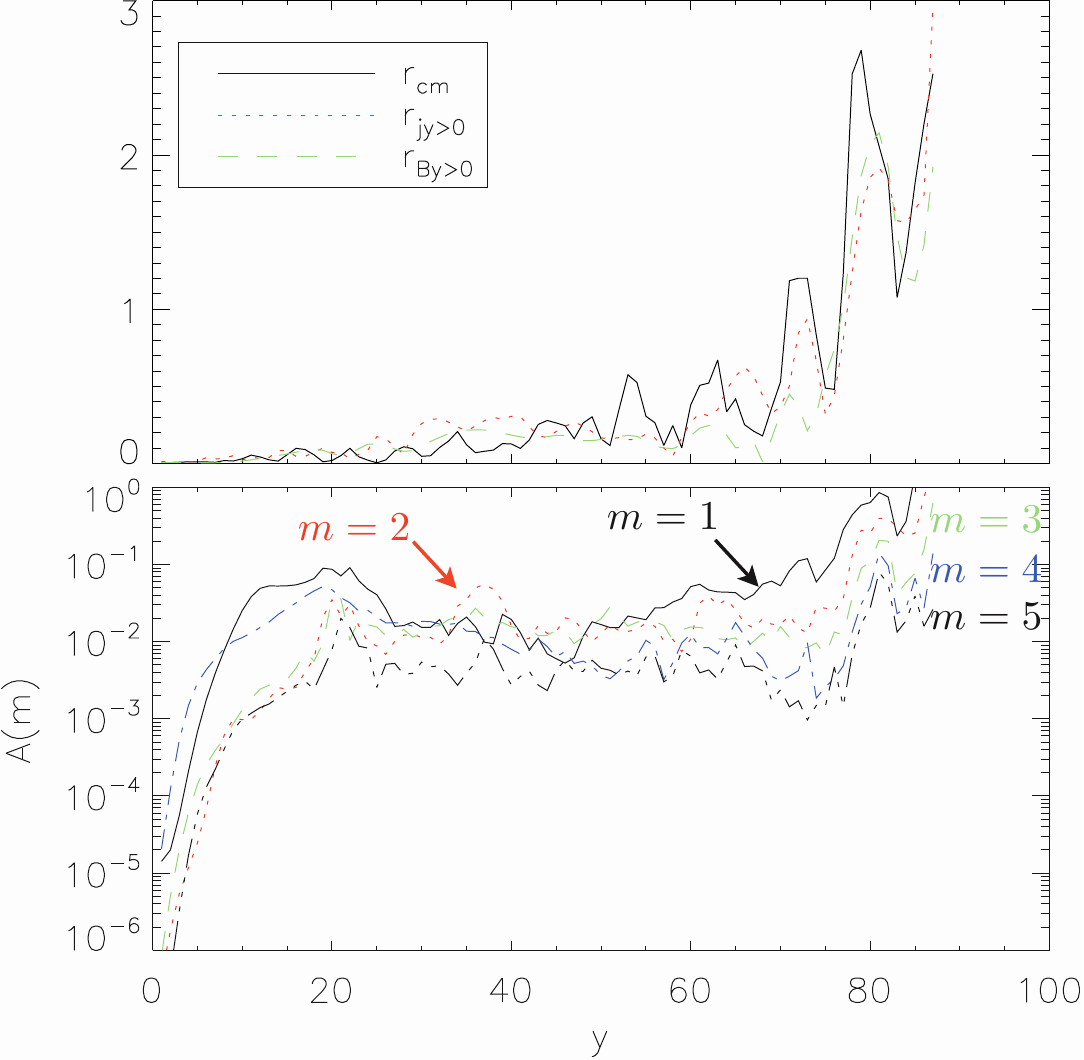}
\caption{As Figure \ref{fig:barycenters}  but for run L3Dm at time $t=160$.  The mode injection at the base (y=0) avoids artificial $m=4$ dominance.  }
\label{fig:barycentersmodes}
\end{center}
\end{figure}

The clearest indicator of the kink instability can be observed in the deflection of the jet barycenter.  
For this purpose we define the barycenter $\bar{r}=\sqrt{\bar{x}^{2}+\bar{z}^{2}}$ of the quantity Q 
\begin{align}
\bar	{x}\equiv \frac{\int x\ Q\ dx\ dz}{\int Q\ dx\ dz};\hspace{1cm}
\bar	{z}\equiv \frac{\int z\ Q\ dx\ dz}{\int Q\ dx\ dz} \label{eq:barycenters}
\end{align}
in analogy with \cite{Mignone:2010a}.  For the density-displacement $r_{\rm cm}$, we define $Q_{\rm cm}=\chi \Gamma \rho$, where the $\chi$ is computed from the tracer scalars to pick out the jet contribution alone.  This quantity is also shown against simulation time in the right panel of figure \ref{fig:ampvstime}.  For the current displacement $r_{jy>0}$ we set $Q_{jy>0}=\chi j^{+}_{y}$ taking only the positive values of the current density $j^{+}_{y}$. Finally, the motion of the magnetic flux is defined via $Q_{By>0}=\chi B_{y}^{+}$, taking into account only the positive flux $B_{y}^{+}$.  
Instantaneous barycenter position and mode population along the jet is shown in Figure \ref{fig:barycenters} for run L3D and in Figure \ref{fig:barycentersmodes} for the mode-injected run L3Dm.  
Let's first focus on run L3D.  
The barycenter displacement is small compared to the inner disk radius and even tends to decrease along the jet.  Only near the jet-head, at about $y\simeq 80$ a significant displacement can be observed.  
The modes are dominated by ubiquitous $m=4$ noise, which seems to suppress all other fluctuations starting at height $y=20$.  The $m=4$ dominance prevails all the way to the jet head.  

\begin{figure}
\begin{center}
\includegraphics[width=0.49\textwidth]{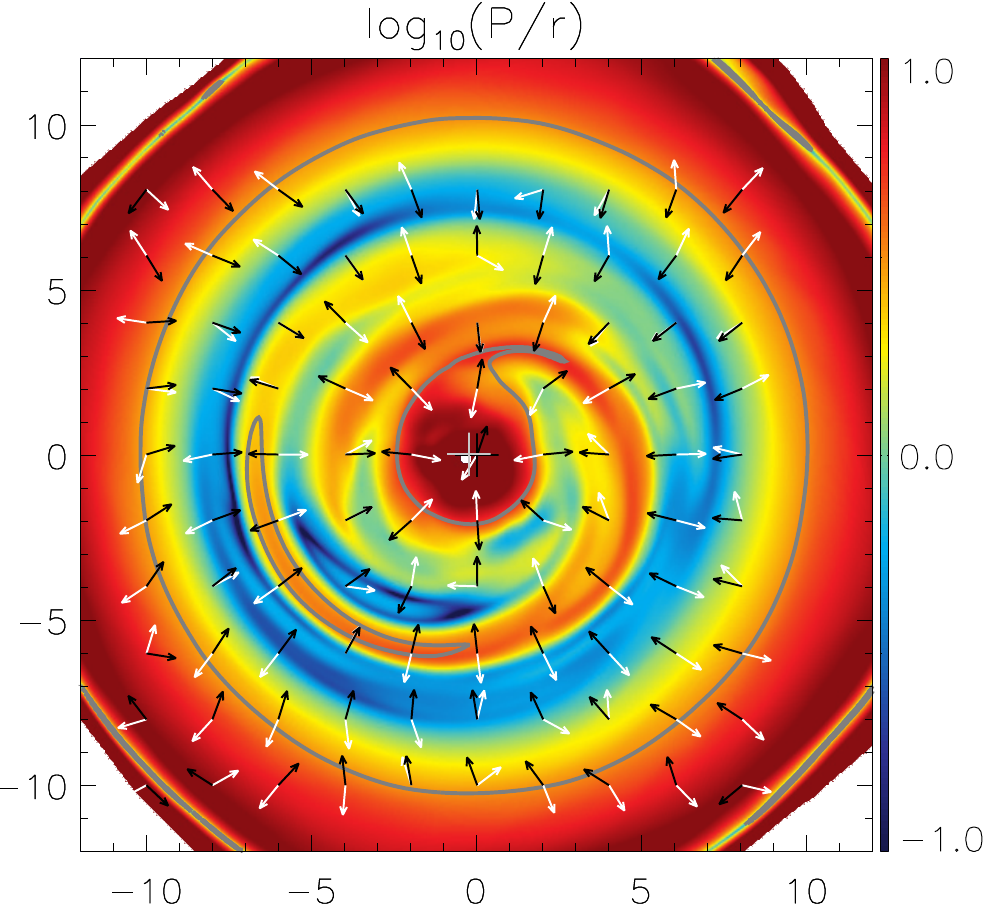}
\includegraphics[width=0.49\textwidth]{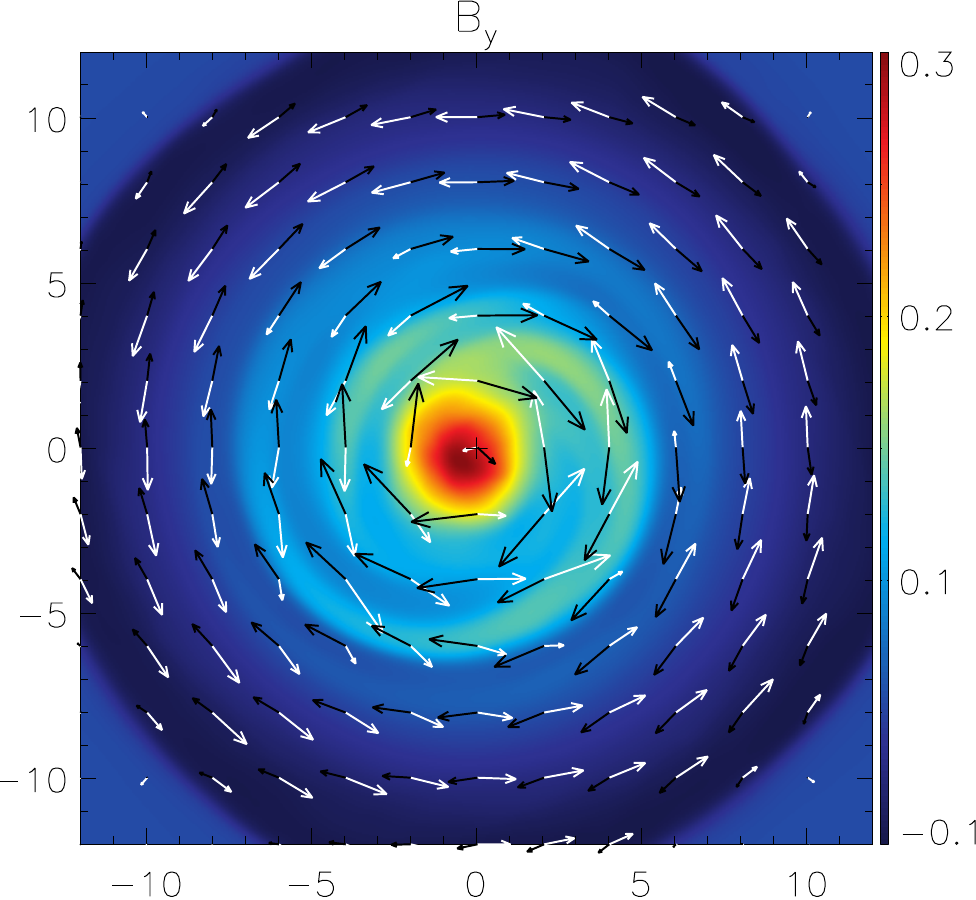}
\hfill
\caption{\textit{Top:}
Co-moving pitch and effective light surface with respect to the center of magnetic flux (shown as grey contours) through the surfaces $y=60$ for run L3Dm at $t=160$.  
Black arrows indicate the direction of the electric force $\rho_{e} \mathbf{E}$ and white arrows show the direction of the Lorentz force $(\nabla \times \mathbf{B}) \times \mathbf{B}$.  Both force vectors are projected onto the image-plane.  
The axis is marked by a black ``+'' and the center of magnetic flux is shown as gray ``+''. 
\text{Bottom:}
Corresponding magnetic flux $B_{y}$ with velocity (white) and magnetic field vectors (black) in the plane.
}
\label{fig:kinkfit}
\end{center}
\end{figure}

The behavior of run L3D is different, here, the kink mode surpasses the $m=4$ at $y\simeq10$.  The angular fluctuations saturate around $y=20$.  
Until $y=60$, the displacements in current and magnetic flux stay roughly constant.  
This hints to a self-stabilization of the jet formation region.  
The kink mode starts to rise again towards the jet head, accompanied with a notable barycenter displacement.    
We note that the magnetic field configuration near the jet head is strongly toroidally dominated as the monopolar initial field configuration is overtaken by the jet (see e.g. Figure \ref{fig:fancyplots}).  
This toroidal dominance could yield an explanation for the stronger growth of the kink mode at the jet head.  
To quantify this further, we introduce the co-moving magnetic pitch defined as 
\begin{align}
P\equiv -2\pi r \frac{B'_{p}}{B'_{\phi}}
\end{align}
which plays a major role in the stability of current carrying plasmas in the laboratory \citep[e.g.][]{Bateman:1978} and astrophysical jets \citep[e.g.][]{appl2000,lery2000}. 
Small values of $P/r<1$ and thus toroidally dominated configurations are particularly susceptible to the kink instability.  
The co-moving fields are obtained by applying the projection $B'^{\alpha} = u_{\beta} F^{*{\alpha \beta}}$, where $F^{*{\alpha \beta}}$ denotes the dual Faraday tensor and $u_{\beta}$ the co-variant Four-velocity as customary.  
The radius $r$ and the fields $B_{\phi}$, $B_{p}$ involved in the definition of the pitch are well-defined however only in axi-symmetry.  
For small perturbations from the cylindrical shape, we can re-orient the symmetry axis to the magnetic backbone at the position $(\bar{x},\bar{z})$ and define effective values for $\bar{r}$, $B_{\bar{\phi}}$ and $B_{\bar{r}}$ with respect to the new origin. To define the magnetic backbone position using equations (\ref{eq:barycenters}), we apply the kernel $Q_{\rm BB}=\chi B_{y}^{2}$ which reliably finds the peak of magnetic flux.  
To investigate also effects due to the electric field, we define an effective light surface via the comparison of the field strengths 
\begin{align}
x=\frac{E}{B_{\bar{p}}} \label{eq:lceff}
\end{align}
where $x=1$ marks the light surface and $B_{\bar{p}}=\sqrt{B_{\bar{r}}^{2}+B_{y}^{2}}\simeq B_{y}$ is the poloidal field strength with respect to the magnetic backbone.  
Since the jet is well collimated $B_{y}\gg B_{\bar{r}}$, the location of the backbone is in fact only of secondary importance for the definition of the light surface and we obtain similar results when considering only the vertical field $B_{y}$ in Equation (\ref{eq:lceff}).  This light surface location is indicated with the gray contour in the top panel of figure \ref{fig:kinkfit} where the pitch profiles across the jet is shown.  
To visualize the Lorentz force across the flow, we show force vectors of the electric field $\rho_{e}\mathbf{E}=(\boldsymbol{\nabla}\cdot \mathbf{E)\ E}$ and of the classical Lorentz force $\mathbf{j\times B}$ where we neglected the displacement-current for simplicity.  

The axial flow is outside of the light surface, then follows a region of super-luminal field line rotation and a third region outside of the main jet where field lines rotate again sub-luminal.  The magnetic backbone is markedly seen in the pitch and we note that electric forces exhibit a de-collimating contribution at the central core while they tend to collimate near the outer light cylinder.  In most regions, the additional electric component is opposed to the classical Lorentz force.  
As a general trend, we obtain a radially decreasing pitch profile from the backbone to its minimal value $P/r\lesssim1$, from where the pitch increases again to the boundary of the jet.  Due to the m=1 motion, a spiral pattern with increased pitch is shed of the magnetic backbone.  This 'smearing' leads to a diminishing of the toroidally dominated regions.  
These regions of low pitch coincide with the locations of super-luminal field-line rotation and the stabilizing effect of electric forces becomes apparent as they tend to counter-act the magnetic contribution.

\subsection{Jet-Cloud interaction}

\begin{figure}
\begin{center}
\includegraphics[width=0.49\textwidth]{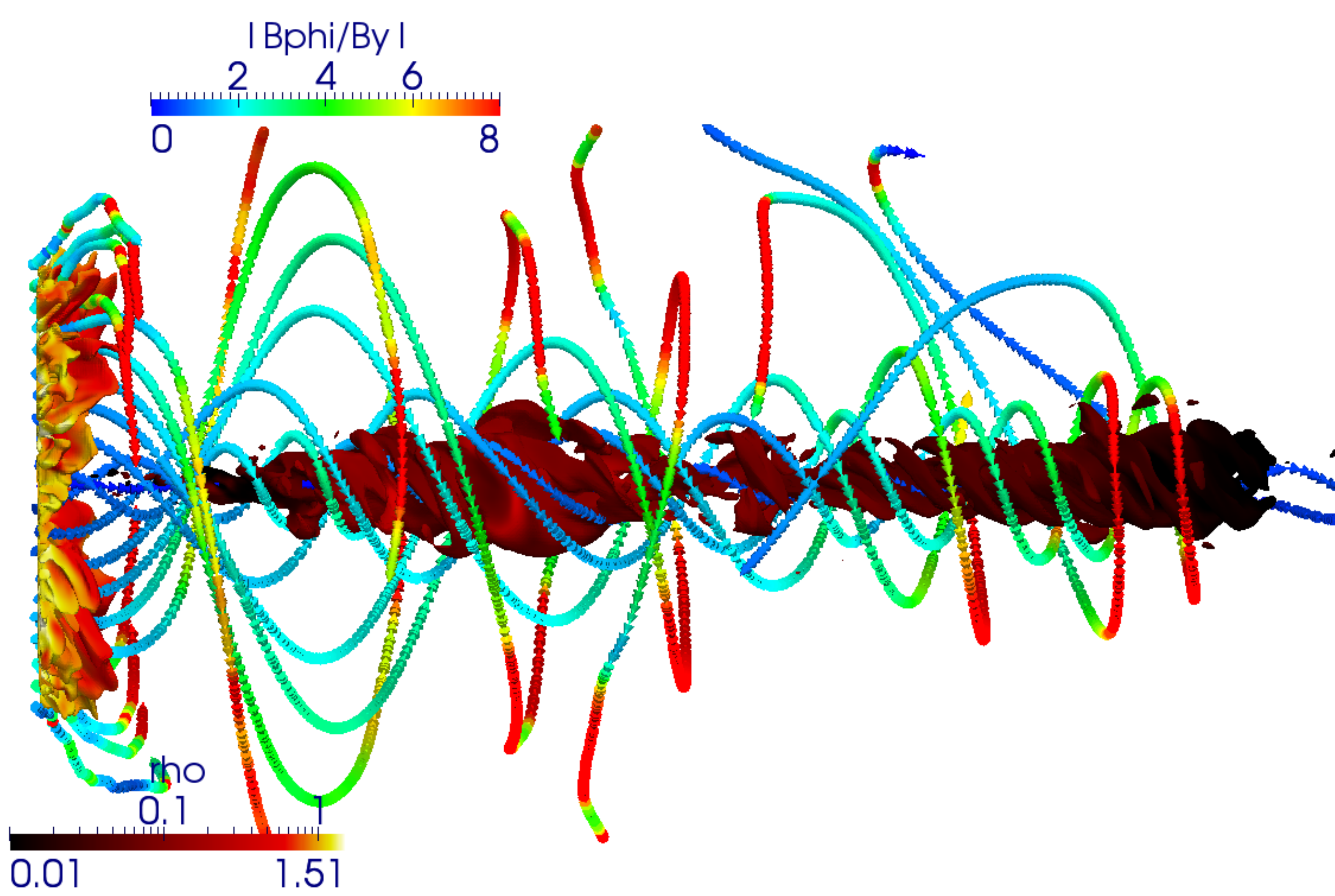}
\includegraphics[width=0.49\textwidth]{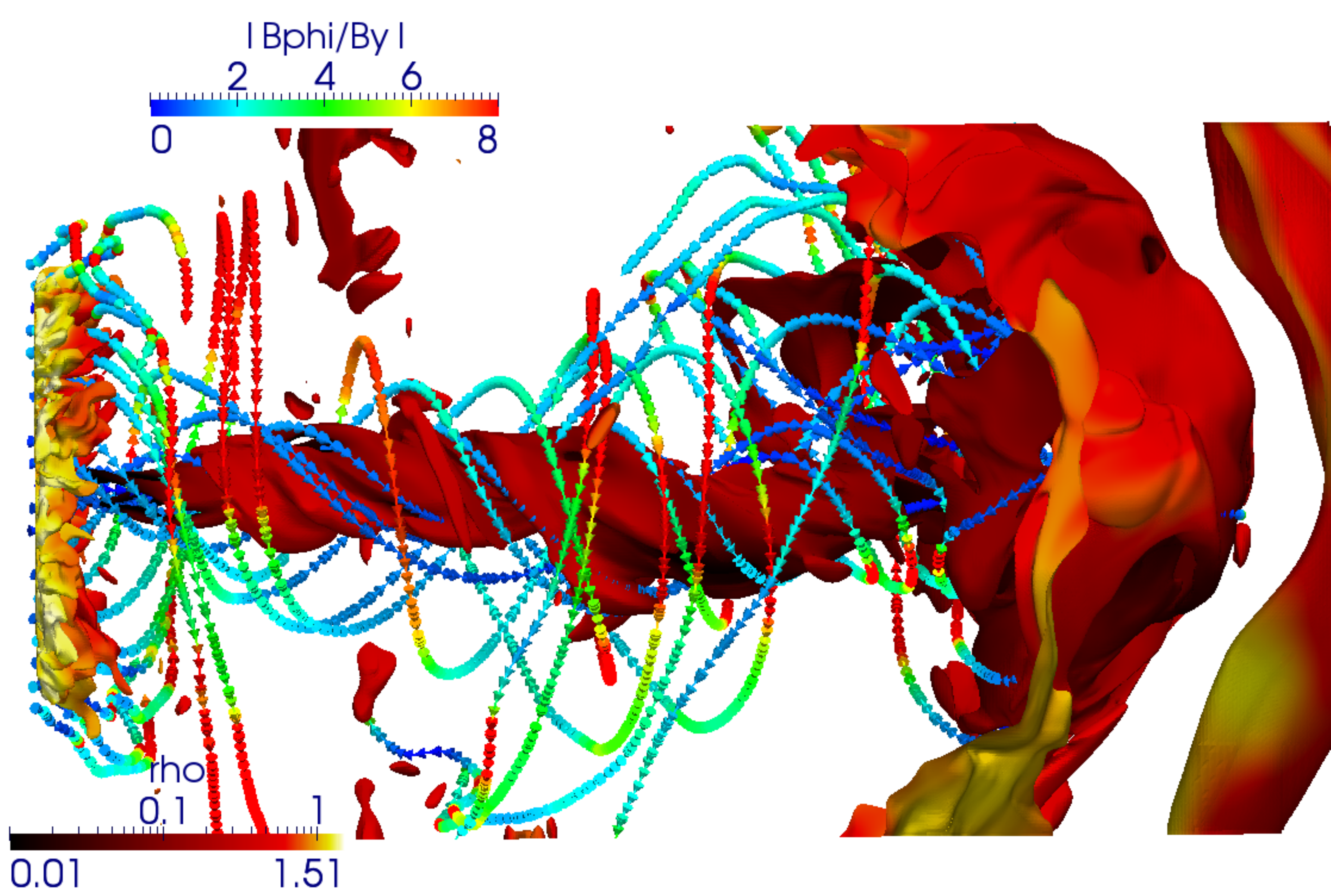}
\caption{Field lines of the unperturbed \textit{(top)} and perturbed \textit{(bottom)} simulations at times $t=103$ and $t=194$, respectively.  To guide the eye along the bent jet, pressure isocontours are added in the figures.  The perturbed run shows a wider magnetic backbone and decreased toroidal field.  }
\label{fig:cloudplot}
\end{center}
\end{figure}

\begin{figure}
\begin{center}
\includegraphics[width=0.46\textwidth]{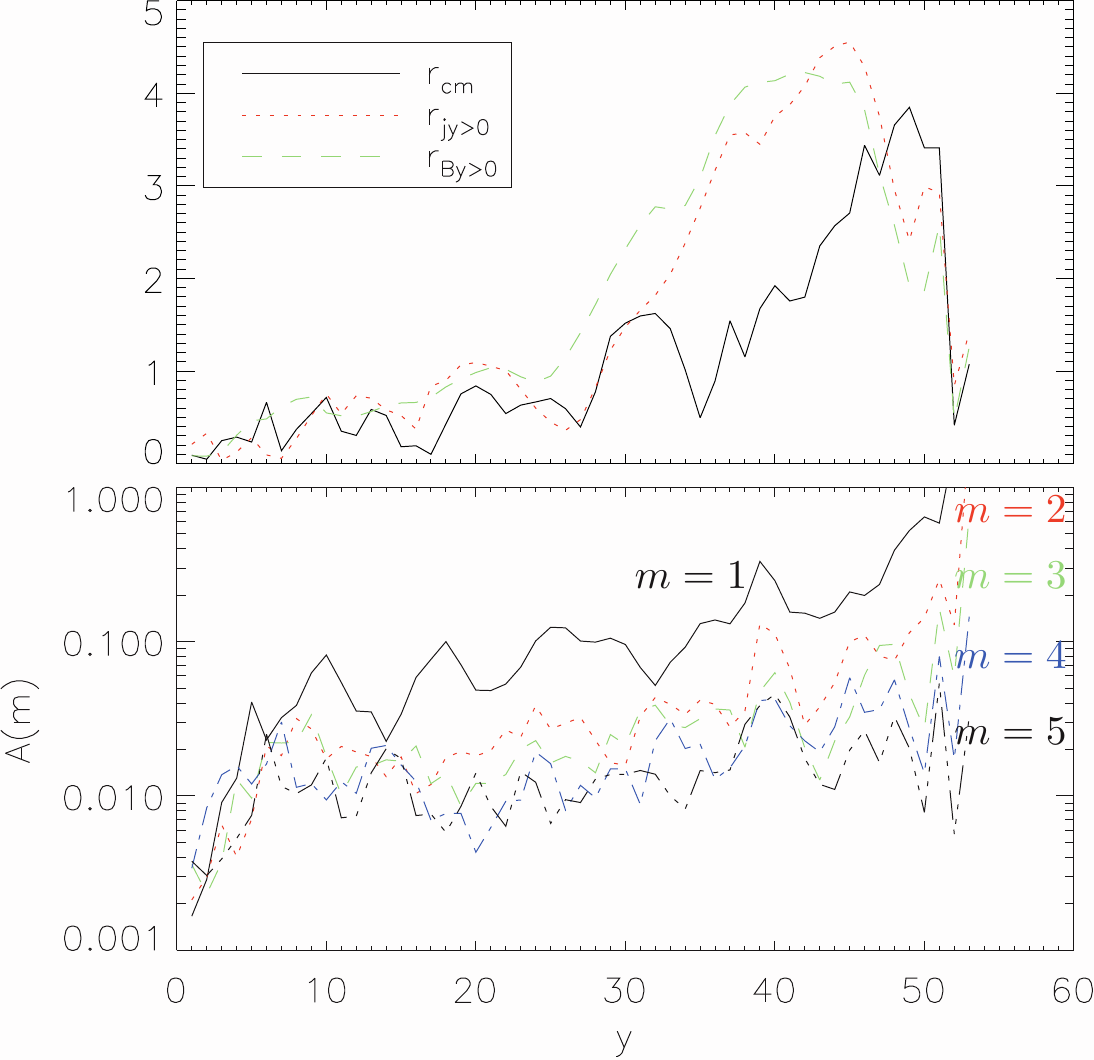}
\hfill
\caption{Barycenter motion and low order angular modes in density of simulation M3Dmd at $t=194$.  } 
\label{fig:barycl}
\end{center}
\end{figure}

\begin{figure}
\begin{center}
\includegraphics[width=0.23\textwidth]{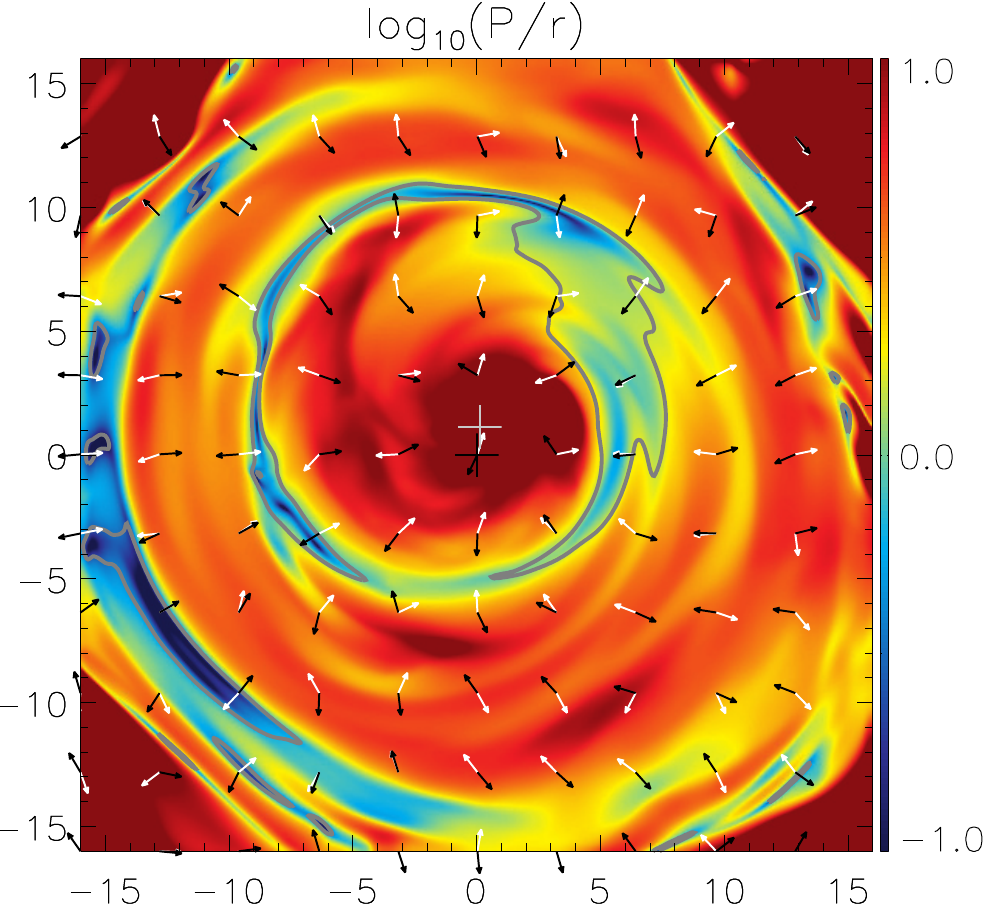}
\includegraphics[width=0.23\textwidth]{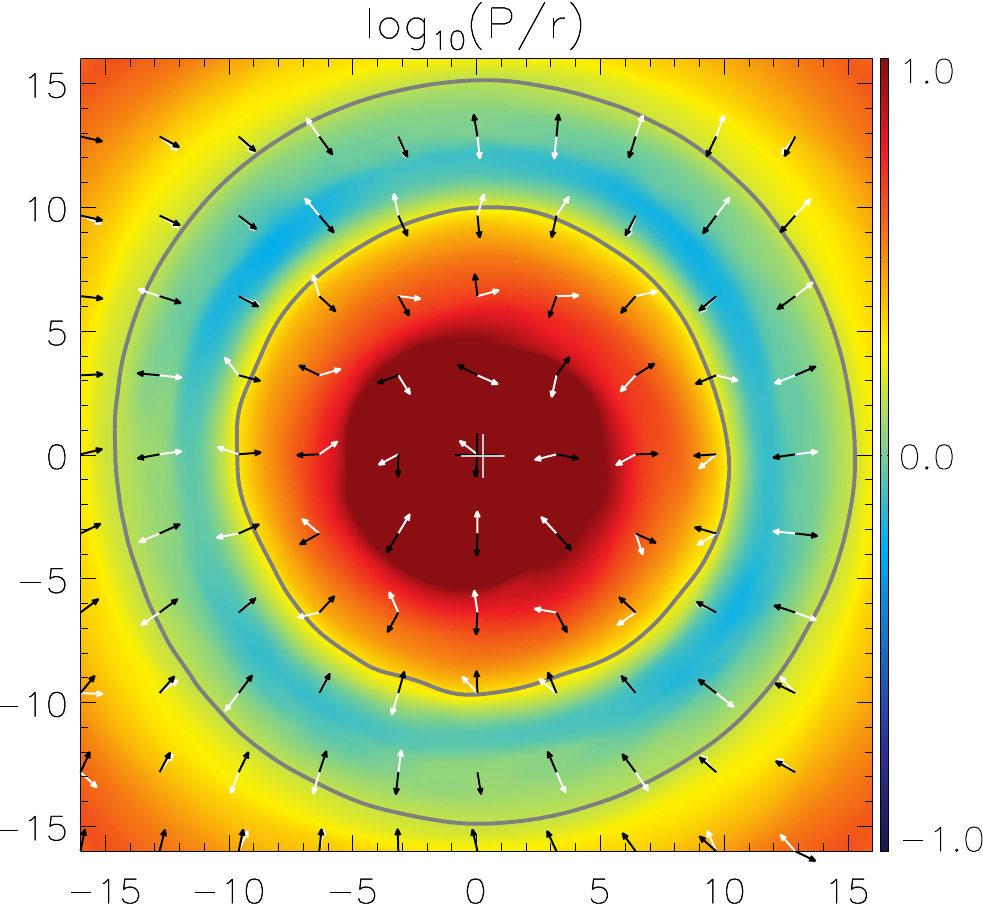}
\hfill
\caption{As in Figure \ref{fig:kinkfit}; comparison of the pitch at $y=32$ in runs M3Dmd \textit{(left)} and L3Dm \textit{(right)} at times $t=194$ respectively $t=160$.  The motion of the backbone in the heavily perturbed simulation increases the poloidal field \new{on account of} the unwinding of the helical field.  }
\label{fig:pitchclouds}
\end{center}
\end{figure}

We now study how the jet reacts to external perturbations exerted by a clumpy ambient medium.  As the jet head punches its way through the in-homogeneous environment, also the upstream flow becomes deflected.  
Figure \ref{fig:cloudplot} shows a rendering of the simulation M3Dmd with mode injection and clumpy medium in comparison to the unperturbed case M3D.  Field lines are colored according to the magnetic pitch.  The clouds with a density contrast of 100:1 represent a strong perturbation and we find the jet heavily distorted.  
Accordingly, the motion of the jet barycenter is increased and we find a strong dominance of the $m=1$ mode along the whole jet (Figure \ref{fig:barycl}).  
Due to the increased density of the external medium, the jet propagation speed is reduced and the bow-shock is much wider (compare with Figure \ref{fig:fancyplots}).  We therefore compare the jet morphology between runs M3D and M3Dmd at times of roughly equal jet propagation length.  
The precession of the magnetic backbone against the toroidal magnetic field direction tends to ``smear'' out the high-pitched axial region and the tightly wound helix is effectively unwound.  
This represents an efficient mechanism of jet self-stabilization also noted by \cite{2003ApJ...582..292O}.  In their study the effect was described as follows: ``The appearance of the $|m|=1$ modes pumps energy into the poloidal magnetic field, causing the jet Alfv\'en Mach number to fall below unity and stabilize the jet''.  From our simulations we come to a similar conclusion, in addition, we note that the ``unwinding'' of the helical field also decreases the field-line rotation $\Omega$ and thus also the influence of electric fields.  We show the increase in magnetic pitch compared to the unperturbed simulation L3D in Figure \ref{fig:pitchclouds}.  
Regions of low pitch are reduced to a thin sheet where the effective light cylinder is found.  In this regime, electric stabilization can not be of importance since the typical value of the light-cylinder $x$ is only of order unity or below.  

\begin{figure}
\begin{center}
\includegraphics[width=0.45\textwidth]{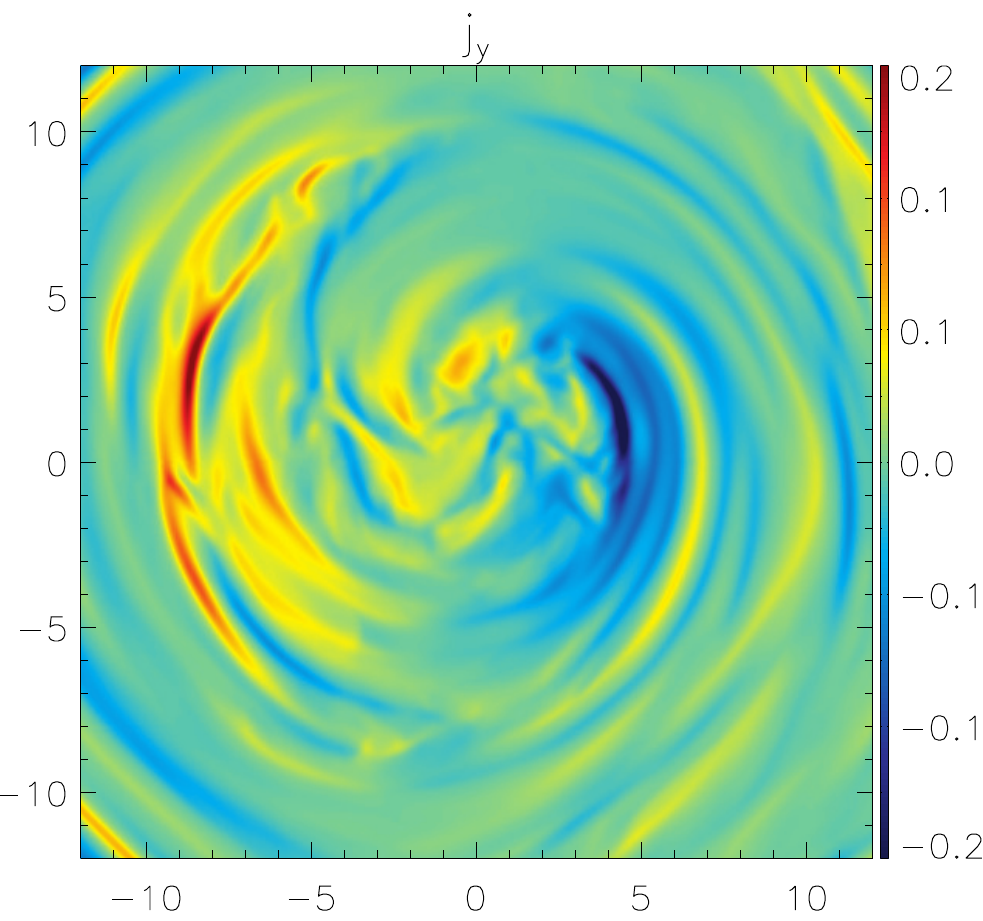}
\hfill
\caption{Out of plane current density at $y=32$, $t=194$ for the run with external perturbation M3Dmd. A layered filamentary structure of alternating current directions develops.  }
\label{fig:jyclouds}
\end{center}
\end{figure}

It is interesting to visualize the return current in the heavily perturbed run M3Dmd.  
The structure of the current density across the jet is shown in figure \ref{fig:jyclouds}.  Fine layered current sheets form on small scales that could in principle dissipate and give rise to particle acceleration within the jet.  It is only due to the high resolution that this effect can be observed.

\section{Conclusions}

We have presented first results of high-resolution 3D simulations of relativistic jet formation from magnetospheres in Keplerian rotation.  
When the flow is perturbed by non-axisymmetric internal perturbations of the accretion disk corona, the modes first grow exponentially at the base, approach saturation along the jet and grow again towards the jet head.  The $m=1$ kink is the dominant mode of departure from axisymmetry.  At a given height above the accretion disk, the temporal evolution of the modes was considered.  Also here we find a saturation of perturbations before a notable dissipation or even disruption is encountered.  
As an aside, we also performed simulations where the only measure of perturbation is the ubiquitous discretization noise of the grid.  In result, the modes are dominated by multiples of $m=4$ which grow along the flow at \new{the expense of} other modes and we observe virtually no motion of the jet barycenter.  

To further investigate the stability of the jet structure, we considered the co-moving magnetic pitch.  As in the axisymmetric case, a stabilizing backbone of high pitch $P\gg1$ develops, surrounded by an intermediate, toroidally dominated region $P<1$ and an outer high-pitch region at the border of the jet.  The locations of super-luminal field line rotation (the light surface) approximately coincide with the low-pitch region.  
Forces due to the electric field $\rho_{e}\mathbf{E}$ oppose the classical magnetic Lorentz force $\mathbf{(\boldsymbol{\nabla}\times B) \times B}$ which could thus add to jet stabilization in the relativistic case.  

In order to study external perturbations, we initialized the domain with a static clumpy ambient medium following a power-law spectrum in Fourier space.  
While this should not be mistaken for a fully developed MHD-turbulent medium, it allows us to investigate perturbations ``external'' to the jet as a general scenario.  
The maximum amplitude in cloud density to the jet density was chosen as $100:1$ which thus represent a strong perturbation to the emerging jet.  

In result, the jet funnels its way through the path of least resistance which leads to large departures of the barycenter from the axis and dominating $m=1$ modes.  
When compared to the unperturbed case, the magnetic pitch is largely increased which can be interpreted as a mechanism of jet self-stabilization.  Due to the precession of the magnetic backbone against the toroidal magnetic field direction, the helical structure tends to be ``unwound'' leading to an increase of poloidal field which also reduces the amount of field-line rotation.  
The external perturbation and accompanying motion of the magnetic backbone also gives rise to filamentary small-scale structure visible in the return currents.  The accompanying dissipation could facilitate particle acceleration within the jet.  
Further investigation with high-resolution 3D simulations is needed to quantify this effect.  
\new{Another point of interest is the validity of the axisymmetric assumption in the formation of jets in general.  We find that due to the early saturation of non-axisymmetric instabilities, 2D results on the inner scales of jet acceleration and collimation (e.g. PF I, PF II) appear largely unaffected by non-axisymmetric effects.  Detailed comparisons of the dynamics in two- and three- dimensions will be provided in a forthcoming paper.}

\section*{Acknowledgments}
The author likes to thank Serguei Komissarov for comments on the early manuscript and Rony Keppens for remarks on the later version.   Sincere thanks go to Christian Fendt for invaluable advice and many discussions on the formation of jets.  
Computing for this work was carried out under the HPC-EUROPA2 project (project number: 228398), with the support of the European Community - Research Infrastructure Action of the FP7.
Post-processing of the simulations was performed on the VIZ cluster of the Max Planck Society.

\bibliographystyle{mn2e} 
\bibliography{astro}

\begin{thebibliography}{}

\bibitem[\protect\citeauthoryear{{Anderson}, {Li}, {Krasnopolsky} \&
  {Blandford}}{{Anderson} et~al.}{2006}]{anderson2006}
{Anderson} J.~M.,  {Li} Z.,  {Krasnopolsky} R.,    {Blandford} R.~D.,  2006,
  \apjl, 653, L33

\bibitem[\protect\citeauthoryear{{Appl} \& {Camenzind}}{{Appl} \&
  {Camenzind}}{1992}]{appl1992}
{Appl} S.,  {Camenzind} M.,  1992, \aap, 256, 354

\bibitem[\protect\citeauthoryear{{Appl}, {Lery} \& {Baty}}{{Appl}
  et~al.}{2000}]{appl2000}
{Appl} S.,  {Lery} T.,    {Baty} H.,  2000, \aap, 355, 818

\bibitem[\protect\citeauthoryear{{Araudo}, {Bosch-Ramon} \& {Romero}}{{Araudo}
  et~al.}{2010}]{araudo2010}
{Araudo} A.~T.,  {Bosch-Ramon} V.,    {Romero} G.~E.,  2010, \aap, 522, A97+

\bibitem[\protect\citeauthoryear{{Bateman}}{{Bateman}}{1978}]{Bateman:1978}
{Bateman} G.,  1978, {MHD instabilities}.
MIT Press

\bibitem[\protect\citeauthoryear{{Begelman}}{{Begelman}}{1998}]{begelman1998}
{Begelman} M.~C.,  1998, \apj, 493, 291

\bibitem[\protect\citeauthoryear{{Blandford} \& {K\"onigl}}{{Blandford} \&
  {K\"onigl}}{1979}]{Blandford1979}
{Blandford} R.~D.,  {K\"onigl} A.,  1979, \apj, 232, 34

\bibitem[\protect\citeauthoryear{{Blandford} \& {Znajek}}{{Blandford} \&
  {Znajek}}{1977}]{1977MNRAS.179..433B}
{Blandford} R.~D.,  {Znajek} R.~L.,  1977, \mnras, 179, 433

\bibitem[\protect\citeauthoryear{{B{\"o}ttcher}}{{B{\"o}ttcher}}{2007}]{bottcher2007}
{B{\"o}ttcher} M.,  2007, \apss, 309, 95

\bibitem[\protect\citeauthoryear{{Camenzind}}{{Camenzind}}{1986}]{Camenzind1986}
{Camenzind} M.,  1986, \aap, 162, 32

\bibitem[\protect\citeauthoryear{{Davidson} \& {Netzer}}{{Davidson} \&
  {Netzer}}{1979}]{davidson1979}
{Davidson} K.,  {Netzer} H.,  1979, Reviews of Modern Physics, 51, 715

\bibitem[\protect\citeauthoryear{{Drenkhahn} \& {Spruit}}{{Drenkhahn} \&
  {Spruit}}{2002}]{2002A&A...391.1141D}
{Drenkhahn} G.,  {Spruit} H.~C.,  2002, \aap, 391, 1141

\bibitem[\protect\citeauthoryear{{Dullemond} \& {van Bemmel}}{{Dullemond} \&
  {van Bemmel}}{2005}]{dullemond2005}
{Dullemond} C.~P.,  {van Bemmel} I.~M.,  2005, \aap, 436, 47

\bibitem[\protect\citeauthoryear{{Hardee} \& {Hughes}}{{Hardee} \&
  {Hughes}}{2003}]{2003ApJ...583..116H}
{Hardee} P.~E.,  {Hughes} P.~A.,  2003, \apj, 583, 116

\bibitem[\protect\citeauthoryear{{Heinz} \& {Begelman}}{{Heinz} \&
  {Begelman}}{2000}]{heinz2000}
{Heinz} S.,  {Begelman} M.~C.,  2000, \apj, 535, 104

\bibitem[\protect\citeauthoryear{{Istomin} \& {Pariev}}{{Istomin} \&
  {Pariev}}{1994}]{istomin1994}
{Istomin} Y.~N.,  {Pariev} V.~I.,  1994, \mnras, 267, 629

\bibitem[\protect\citeauthoryear{{Istomin} \& {Pariev}}{{Istomin} \&
  {Pariev}}{1996}]{istomin1996}
{Istomin} Y.~N.,  {Pariev} V.~I.,  1996, \mnras, 281, 1

\bibitem[\protect\citeauthoryear{{Jorstad}, {Marscher}, {Lister}, {Stirling},
  {Cawthorne}, {Gear}, {G{\'o}mez}, {Stevens}, {Smith}, {Forster} \&
  {Robson}}{{Jorstad} et~al.}{2005}]{jorstad2005}
{Jorstad} S.~G.,  {Marscher} A.~P.,  {Lister} M.~L.,  {Stirling} A.~M.,
  {Cawthorne} T.~V.,  {Gear} W.~K.,  {G{\'o}mez} J.~L.,  {Stevens} J.~A.,
  {Smith} P.~S.,  {Forster} J.~R.,    {Robson} E.~I.,  2005, \aj, 130, 1418

\bibitem[\protect\citeauthoryear{Keppens, Meliani, van Marle, Delmont, Vlasis
  \& van~der Holst}{Keppens et~al.}{2012}]{Keppens2012718}
Keppens R.,  Meliani Z.,  van Marle A.,  Delmont P.,  Vlasis A.,    van~der
  Holst B.,  2012, Journal of Computational Physics, 231, 718

\bibitem[\protect\citeauthoryear{{Keppens}, {T{\'o}th}, {Westermann} \&
  {Goedbloed}}{{Keppens} et~al.}{1999}]{keppens1999}
{Keppens} R.,  {T{\'o}th} G.,  {Westermann} R.~H.~J.,    {Goedbloed} J.~P.,
  1999, Journal of Plasma Physics, 61, 1

\bibitem[\protect\citeauthoryear{{Komissarov}}{{Komissarov}}{2001}]{komissarov2001}
{Komissarov} S.~S.,  2001, \mnras, 326, L41

\bibitem[\protect\citeauthoryear{{Komissarov}, {Barkov}, {Vlahakis} \&
  {K{\"o}nigl}}{{Komissarov} et~al.}{2007}]{2007MNRAS.380...51K}
{Komissarov} S.~S.,  {Barkov} M.~V.,  {Vlahakis} N.,    {K{\"o}nigl} A.,  2007,
  \mnras, 380, 51

\bibitem[\protect\citeauthoryear{{Lery}, {Baty} \& {Appl}}{{Lery}
  et~al.}{2000}]{lery2000}
{Lery} T.,  {Baty} H.,    {Appl} S.,  2000, \aap, 355, 1201

\bibitem[\protect\citeauthoryear{{Lister}, {Cohen}, {Homan}, {Kadler},
  {Kellermann}, {Kovalev}, {Ros}, {Savolainen} \& {Zensus}}{{Lister}
  et~al.}{2009}]{lister2009}
{Lister} M.~L.,  {Cohen} M.~H.,  {Homan} D.~C.,  {Kadler} M.,  {Kellermann}
  K.~I.,  {Kovalev} Y.~Y.,  {Ros} E.,  {Savolainen} T.,    {Zensus} J.~A.,
  2009, \aj, 138, 1874

\bibitem[\protect\citeauthoryear{{Lohner}}{{Lohner}}{1987}]{lohner1987}
{Lohner} R.,  1987, Computer Methods in Applied Mechanics and Engineering, 61,
  323

\bibitem[\protect\citeauthoryear{{Lyubarskii}}{{Lyubarskii}}{1999}]{Lyubarskii:1999}
{Lyubarskii} Y.~E.,  1999, \mnras, 308, 1006

\bibitem[\protect\citeauthoryear{{Lyutikov} \& {Uzdensky}}{{Lyutikov} \&
  {Uzdensky}}{2003}]{2003ApJ...589..893L}
{Lyutikov} M.,  {Uzdensky} D.,  2003, \apj, 589, 893

\bibitem[\protect\citeauthoryear{{McKinney} \& {Blandford}}{{McKinney} \&
  {Blandford}}{2009}]{mckinney2009}
{McKinney} J.~C.,  {Blandford} R.~D.,  2009, \mnras, 394, L126

\bibitem[\protect\citeauthoryear{{Meliani}, {Sauty}, {Tsinganos} \&
  {Vlahakis}}{{Meliani} et~al.}{2004}]{meliani2004}
{Meliani} Z.,  {Sauty} C.,  {Tsinganos} K.,    {Vlahakis} N.,  2004, \aap, 425,
  773

\bibitem[\protect\citeauthoryear{{Michel}}{{Michel}}{1969}]{1969ApJ...158..727M}
{Michel} F.~C.,  1969, \apj, 158, 727

\bibitem[\protect\citeauthoryear{{Mignone}, {Rossi}, {Bodo}, {Ferrari} \&
  {Massaglia}}{{Mignone} et~al.}{2010}]{Mignone:2010a}
{Mignone} A.,  {Rossi} P.,  {Bodo} G.,  {Ferrari} A.,    {Massaglia} S.,  2010,
  \mnras, 402, 7

\bibitem[\protect\citeauthoryear{{Mizuno}, {Hardee} \& {Nishikawa}}{{Mizuno}
  et~al.}{2007}]{2007ApJ...662..835M}
{Mizuno} Y.,  {Hardee} P.,    {Nishikawa} K.,  2007, \apj, 662, 835

\bibitem[\protect\citeauthoryear{{Mizuno}, {Hardee} \& {Nishikawa}}{{Mizuno}
  et~al.}{2011}]{mizuno2011a}
{Mizuno} Y.,  {Hardee} P.~E.,    {Nishikawa} K.-I.,  2011, \apj, 734, 19

\bibitem[\protect\citeauthoryear{{Mizuno}, {Lyubarsky}, {Nishikawa} \&
  {Hardee}}{{Mizuno} et~al.}{2009}]{2009ApJ...700..684M}
{Mizuno} Y.,  {Lyubarsky} Y.,  {Nishikawa} K.,    {Hardee} P.~E.,  2009, \apj,
  700, 684

\bibitem[\protect\citeauthoryear{{Moll}, {Spruit} \& {Obergaulinger}}{{Moll}
  et~al.}{2008}]{2008A&A...492..621M}
{Moll} R.,  {Spruit} H.~C.,    {Obergaulinger} M.,  2008, \aap, 492, 621

\bibitem[\protect\citeauthoryear{{Narayan}, {Li} \& {Tchekhovskoy}}{{Narayan}
  et~al.}{2009}]{narayan2009}
{Narayan} R.,  {Li} J.,    {Tchekhovskoy} A.,  2009, \apj, 697, 1681

\bibitem[\protect\citeauthoryear{{O'Neill}, {Beckwith} \& {Begelman}}{{O'Neill}
  et~al.}{2012}]{oneill2012a}
{O'Neill} S.~M.,  {Beckwith} K.,    {Begelman} M.~C.,  2012, \mnras, 422, 1436

\bibitem[\protect\citeauthoryear{{Ouyed}, {Clarke} \& {Pudritz}}{{Ouyed}
  et~al.}{2003}]{2003ApJ...582..292O}
{Ouyed} R.,  {Clarke} D.~A.,    {Pudritz} R.~E.,  2003, \apj, 582, 292

\bibitem[\protect\citeauthoryear{{Porth} \& {Fendt}}{{Porth} \&
  {Fendt}}{2010}]{2010ApJ...709.1100P}
{Porth} O.,  {Fendt} C.,  2010, \apj, 709, 1100

\bibitem[\protect\citeauthoryear{{Porth}, {Fendt}, {Meliani} \&
  {Vaidya}}{{Porth} et~al.}{2011}]{porth2011}
{Porth} O.,  {Fendt} C.,  {Meliani} Z.,    {Vaidya} B.,  2011, \apj, 737, 42

\bibitem[\protect\citeauthoryear{{Rosen} \& {Hardee}}{{Rosen} \&
  {Hardee}}{2000}]{2000ApJ...542..750R}
{Rosen} A.,  {Hardee} P.~E.,  2000, \apj, 542, 750

\bibitem[\protect\citeauthoryear{{Rossi}, {Mignone}, {Bodo}, {Massaglia} \&
  {Ferrari}}{{Rossi} et~al.}{2008}]{rossi2008}
{Rossi} P.,  {Mignone} A.,  {Bodo} G.,  {Massaglia} S.,    {Ferrari} A.,  2008,
  \aap, 488, 795

\bibitem[\protect\citeauthoryear{{Staff}, {Niebergal}, {Ouyed}, {Pudritz} \&
  {Cai}}{{Staff} et~al.}{2010}]{staff2010}
{Staff} J.~E.,  {Niebergal} B.~P.,  {Ouyed} R.,  {Pudritz} R.~E.,    {Cai} K.,
  2010, \apj, 722, 1325

\bibitem[\protect\citeauthoryear{Synge}{Synge}{1957}]{synge1957}
Synge J.~L.,  1957, The relativistic gas..
North-Holland Pub. Co

\bibitem[\protect\citeauthoryear{{Tchekhovskoy}, {McKinney} \&
  {Narayan}}{{Tchekhovskoy} et~al.}{2008}]{2008MNRAS.388..551T}
{Tchekhovskoy} A.,  {McKinney} J.~C.,    {Narayan} R.,  2008, \mnras, 388, 551

\bibitem[\protect\citeauthoryear{{Todo}, {Uchida}, {Sato} \& {Rosner}}{{Todo}
  et~al.}{1993}]{todo1993}
{Todo} Y.,  {Uchida} Y.,  {Sato} T.,    {Rosner} R.,  1993, \apj, 403, 164

\bibitem[\protect\citeauthoryear{{Tomimatsu}, {Matsuoka} \&
  {Takahashi}}{{Tomimatsu} et~al.}{2001}]{Tomimatsu:2001}
{Tomimatsu} A.,  {Matsuoka} T.,    {Takahashi} M.,  2001, \prd, 64, 123003

\bibitem[\protect\citeauthoryear{{van der Klis}, {Jansen}, {van Paradijs},
  {Lewin}, {van den Heuvel}, {Trumper} \& {Szatjno}}{{van der Klis}
  et~al.}{1985}]{van-der-klis1985}
{van der Klis} M.,  {Jansen} F.,  {van Paradijs} J.,  {Lewin} W.~H.~G.,  {van
  den Heuvel} E.~P.~J.,  {Trumper} J.~E.,    {Szatjno} M.,  1985, \nat, 316,
  225

\end{thebibliography}

\bsp

\label{lastpage}

\end{document}